\documentclass[11pt,preprint]{aastex}
\usepackage[numberedappendix]{emulateapj5}
\shorttitle{%
  Primordial Non-Gaussianity of CMB and LSS
}%
\shortauthors{%
  Hikage, Komatsu, \& Matsubara
}%
\begin{document}
\title{%
  Primordial Non-Gaussianity and Analytical Formula for \\
  Minkowski Functionals of the Cosmic Microwave Background \\ 
  and Large-scale Structure
}%
\author{%
  Chiaki Hikage\altaffilmark{1}, 
  Eiichiro Komatsu\altaffilmark{2}, 
  and Takahiko Matsubara\altaffilmark{1}
}%
\altaffiltext{1}{%
  Department of Physics and Astrophysics, 
  Nagoya University, Chikusa, Nagoya 464-8602, Japan
}%
\altaffiltext{2}{%
  Department of Astronomy, 
  University of Texas at Austin, 1 University Station,
  C1400, Austin TX 78712
}%
\email{%
  hikage@a.phys.nagoya-u.ac.jp, komatsu@astro.as.utexas.edu, taka@nagoya-u.jp
}%
\submitted{Received 2006 July 12; accepted 2006 August 22}
\begin{abstract}
 We derive analytical formulae for the Minkowski Functions of the
 cosmic microwave background (CMB) and large-scale structure (LSS)
 from primordial non-Gaussianity. These formulae enable us to estimate
 a non-linear coupling parameter, $f_{\rm NL}$, directly from the CMB
 and LSS data without relying on numerical simulations of non-Gaussian
 primordial fluctuations. One can use these formulae to estimate
 statistical errors on $f_{\rm NL}$ from {\it Gaussian} realizations,
 which are much faster to generate than non-Gaussian ones, fully
 taking into account the cosmic/sampling variance, beam smearing,
 survey mask, etc.  We show that the CMB data from the Wilkinson
 Microwave Anisotropy Probe should be sensitive to $|f_{\rm NL}|\simeq
 40$ at the $68\%$ confidence level. The Planck data should be
 sensitive to $|f_{\rm NL}|\simeq 20$. As for the LSS data, the
 late-time non-Gaussianity arising from gravitational instability and
 galaxy biasing makes it more challenging to detect primordial
 non-Gaussianity at low redshifts.  The late-time effects obscure the
 primordial signals at small spatial scales. High-redshift galaxy
 surveys at $z>2$ covering $\sim 10$~Gpc$^3$ volume would be required
 for the LSS data to detect $|f_{\rm NL}|\simeq 100$. Minkowski
 Functionals are nicely complementary to the bispectrum because the
 Minkowski Functionals are defined in real space and the bispectrum is
 defined in Fourier space. This property makes the Minkowski
 Functionals a useful tool in the presence of real-world issues such
 as anisotropic noise, foreground and survey masks. Our formalism can
 be extended to scale-dependent $f_{\rm NL}$ easily.
\end{abstract}

\section{Introduction}\label{sec:intro}
Recent observations of cosmological fluctuations from the Cosmic
Microwave Background (CMB) and Large Scale Structure (LSS) strongly
support basic predictions of inflationary scenarios: primordial
fluctuations are nearly scale-invariant
\citep{Spergel2003,Tegmark2004,Seljak2005,Spergel2006}, adiabatic 
\citep{peiris2003,Bucher2004,bean/etal:2006}, 
and Gaussian \citep[][and references
therein]{Komatsu2002,Komatsu2003,Creminelli2005,Spergel2006}.  In
order to discriminate between more-than-100 candidate inflationary
models, however, one needs to look for {\it deviations} from the
scale-invariance, adiabaticity as well as Gaussianity, for which
different inflationary models make specific predictions.  Inflationary
models based upon a slowly rolling single-field scalar field generally
predict very small deviations from Gaussianity; however, the
post-inflationary evolution of non-linear metric perturbations
inevitably generates ubiquitous non-Gaussian fluctuations.  On the other
hand, a broad class of inflationary models based upon different
assumptions about the nature of scalar field(s) can generate significant
primordial non-Gaussianity
\citep{lyth/etal:2003,dvali/etal:2004,arkani-hamed/etal:2004,alishahiha/etal:2004,BKMR2004}. Therefore,
Gaussianity of the primordial fluctuations offers a direct test of
inflationary models. 

It is customary to adopt the following simple form of primordial
non-Gaussianity in Bardeen's curvature perturbations during the matter
era \citep[e.g., ][]{KS2001}: 
\begin{equation}
\label{eq:ngpotential2}
\Phi=\phi+f_{\rm NL}(\phi^2-\langle\phi^2\rangle),
\end{equation}
where $\phi$ is an auxiliary random-Gaussian field and $f_{\rm NL}$
characterizes the amplitude of a quadratic correction to the curvature
perturbations. Note that $\Phi$ is related to the primordial comoving
curvature perturbations generated during inflation, ${\cal R}$, by
$\Phi=(3/5){\cal R}$. While this quadratic form is motivated by
inflationary models based upon a single slowly-rolling scalar field, the
actual predictions usually include momentum dependence in $f_{\rm
NL}$. (That is to say, $f_{\rm NL}$ is not a constant.) Therefore, when
precision is required, one should use the actual formula given by
specific processes, either from primordial non-Gaussianity from
inflation or the post-inflationary evolution of non-linear
perturbations, to calculate a more accurate form of statistical
quantities such as the angular bispectrum of CMB
\citep{Babich2004,Liguori2006}. Nevertheless, a constant $f_{\rm NL}$ is
still a useful parameterization of non-Gaussianity which enables us to
obtain simple analytical formulae for the statistical quantities to
compare with observations. The use of a constant $f_{\rm NL}$ is also
justified by the fact that the current observations are not sensitive
enough to detect momentum-dependence of $f_{\rm NL}$. Therefore, we
adopt the constant $f_{\rm NL}$ for our analysis throughout this
paper. Note that it is actually straightforward to extend our formalism
to a momentum-dependent $f_{\rm NL}$. 

So far, analytical formulae for the statistical quantities of the CMB
from primordial non-Gaussianity are known only for the angular
bispectrum \citep{KS2001,BZ2004,Liguori2006} and trispectrum
\citep{OH2002,KK2006}. The analytical formulae are extremely valuable
especially when one tries to measure non-Gaussian signals from the data.
Fast, nearly optimal estimators for $f_{\rm NL}$ have been derived on
the basis of these analytical formulae \citep{KSW2005,Creminelli2005},
and have been successfully applied to the CMB data from the Wilkinson
Microwave Anisotropy Probe (WMAP): the current constraint on $f_{\rm
NL}$ from the angular bispectrum is $-54$ to $114$ at the 95\%
confidence level \citep{Komatsu2003,Spergel2006}.  \citep[See][for an
alternative parameterization of $f_{\rm NL}$.]{Creminelli2005} As for
the LSS, the analytical formula is known only for the 3-d bispectrum
\citep{Verde2000,Scocci2004}.  The LSS bispectrum contains not only the
primordial non-Gaussianity, but also the late-time non-Gaussianity from
gravitational instability and galaxy biasing, which potentially obscure
the primordial signatures. 

In this paper, we derive analytical formulae for another statistical
tool, namely the Minkowski Functionals (MFs), which describe
morphological properties of fluctuating fields
\citep{MBW1994,SB1997,SG1998,WK98}. In $d$-dimensional space ($d=2$ for
CMB and $d=3$ for LSS), $d+1$ MFs are defined, as listed in
Table~\ref{tab:MFs}. The ``Euler characteristic'' measures topology of
the fields, and is essentially given by the number of hot spots minus
the number of cold spots when $d=2$. This quantity is sometimes called
the ``genus statistics'', which was independently re-discovered by
\citet{GMD1986} in search of a topological measure of non-Gaussianity in
the cosmic density fields. (The Euler characteristic and genus are
different only by a numerical coefficient, $-1/2$.)

\begin{table*}[t]
\caption{Minkowski Functionals defined in $d-$dimensional space: $d=2$
for CMB and $d=3$ for LSS.}
\begin{center}
\begin{tabular}{cccccc}
  \hline\hline
observations & $d$ & $V_0$ & $V_1$ & $V_2$ & $V_3$ \\ 
  \hline
CMB& 2 & Area & Total Circumference & Euler Characteristic & -- \\
LSS& 3 & Volume & Surface Area & Total Mean Curvature & Euler Characteristic \\ \hline
\end{tabular}
\end{center}
\label{tab:MFs}
\end{table*}
Why study MFs? Since different statistical methods are sensitive to
different aspects of non-Gaussianity, one should study as many
statistical methods as possible. Most importantly, the MFs and
bispectrum are very different in that MFs are defined in real space,
whereas the bispectrum is defined in Fourier (or harmonic)
space. Therefore, these statistical methods are nicely complementary
to each other. Previously there are several attempts to give
constraints on the primordial non-Gaussianity using MFs
\citep[e.g,][]{NSM2000}. Although we shall show in this paper that the
MFs do not contain information more than the bispectrum in the limit
that non-Gaussianity is weak, the complementarity is still powerful in
the presence of complicated real-world issues such as inhomogeneous
noise, survey mask, foreground contamination, etc. The MFs have also
been used to constrain $f_{\rm NL}$. \citet{Komatsu2003} and
\citet{Spergel2006} have used numerical simulations of non-Gaussian
CMB sky maps to calculate the predicted form of MFs as a function of
$f_{\rm NL}$, and compared the predictions with the WMAP data to
constrain $f_{\rm NL}$, obtaining similar constraints to the
bispectrum ones. This method (calculating the form of MFs from
non-Gaussian simulations) is, however, a painstaking process: it takes
about three hours to simulate one non-Gaussian map on one processor of
SGI Origin 300.  When cosmological parameters are varied, one needs to
re-simulate a whole batch of non-Gaussian maps from the beginning ---
this is a highly inefficient approach.  Once we have the {\it
analytical} formula for the MFs as a function of $f_{\rm NL}$,
however, we no longer need to simulate non-Gaussian maps, greatly
speeding up the measurement of $f_{\rm NL}$ from the data.

We use the perturbative formula for MFs originally derived by
\citet{Matsubara1994,Matsubara2003}: assuming that non-Gaussianity is
weak, which has been justified by the current constraints on $f_{\rm
NL}$, we consider the lowest-order corrections to the MFs using the
multi-dimensional Edgeworth expansion around a Gaussian distribution
function.

The organization of paper is as follows; In \S~\ref{sec:pb_general} we
review the generic perturbative formula for the Minkowski Functionals. 
In \S~\ref{sec:mf_cmb} we derive the analytical formula for MFs of the
CMB from primordial non-Gaussian fluctuations parameterized by $f_{\rm
NL}$. We also estimate projected statistical errors on $f_{\rm NL}$
expected from the WMAP data from multi-year observations as well as
from the Planck data. In \S~\ref{sec:mf_lss} we derive the analytical
formula for MFs of the LSS from primordial non-Gaussianity, non-linear
gravitational evolution, and galaxy biasing in a perturbative
manner. \S~\ref{sec:summary} is devoted to summary and conclusions. In
Appendix~\ref{app:sim} we outline our method for computing the MFs
from the CMB and LSS data. We also describe our simulations of CMB and
LSS. In Appendix~\ref{app:Bg} we derive the analytical formula for the
galaxy bispectrum. In Appendix~\ref{app:cmbsim} we compare the
analytical MFs of CMB with non-Gaussian simulations in the
Sachs--Wolfe limit. In Appendix~\ref{app:poten}, we extend the
corrections of primordial potential to $n$-th order, in order to
examine more carefully validity of our perturbative expansion.

Throughout this paper, we assume $\Lambda$CDM cosmology with
the cosmological parameters at the maximum likelihood peak 
from the WMAP first-year data only fit
\citep{Spergel2003}. Specifically, we adopt $\Omega_b=0.049$,
$\Omega_{cdm}=0.271$, $\Omega_\Lambda=0.68$, $H_0=68.2~{\rm
km~s^{-1}~Mpc^{-1}}$, $\tau=0.0987$, and $n_s=0.967$. The amplitude of
primordial fluctuations has been normalized by the first acoustic peak
of the temperature power spectrum, $l(l+1)C_l/(2\pi)=(74.7~{\mu}{\rm
K})^2$ at $l=220$ \citep{WMAPpeak}. 

\section{General Perturbative Formula for Minkowski Functionals}
\label{sec:pb_general}
Suppose that we have a $d$-dimensional fluctuating field, $f$, which has
zero mean. Then, one may define the MFs for a given threshold,
$\nu\equiv f/\sigma_0$, where $\sigma_0\equiv \langle f^2\rangle^{1/2}$
is the standard deviation of $f$. \citet{Matsubara2003} has obtained the
analytical formulae for the $k$-th Minkowski Functionals of weakly
non-Gaussian fields in $d$-dimension, $V_k^{(d)}(\nu)$, as
\citep[Eq.~(133) of][]{Matsubara2003} 
\begin{eqnarray}
\label{eq:MFs_perturb}
\nonumber
V_k^{(d)}(\nu) 
& = & \frac1{(2\pi)^{(k+1)/2}}\frac{\omega_d}{\omega_{d-k}\omega_k}
\left(\frac{\sigma_1}{\sqrt{d}\sigma_0}\right)^ke^{-\nu^2/2}\left\{H_{k-1}(\nu)\right.\\   
\nonumber 
& &+
\left[\frac16S^{(0)}H_{k+2}(\nu)
+\frac{k}3S^{(1)}H_k(\nu)\right.\\
& &\left.\left.+\frac{k(k-1)}6S^{(2)}H_{k-2}(\nu)\right]\sigma_0+
{\cal O}(\sigma_0^2)\right\},
\end{eqnarray}
where $H_n(\nu)$ are the Hermite polynomials, and $\omega_k\equiv
\pi^{k/2}/{\Gamma(k/2+1)}$ gives $\omega_0=1$, $\omega_1=2$,
$\omega_2=\pi$, and $\omega_3=4\pi/3$. Here, $S^{(i)}$ are the
``skewness parameters'' defined by 
\begin{eqnarray}
S^{(0)}&\equiv& \frac{\langle f^3\rangle}{\sigma_0^4},\\
S^{(1)}&\equiv& -\frac34\frac{\langle
 f^2(\nabla^2f)\rangle}{\sigma_0^2\sigma_1^2},\\ 
S^{(2)}&\equiv& -\frac{3d}{2(d-1)}\frac{\langle (\nabla f)\cdot(\nabla
 f)(\nabla^2f)\rangle}{\sigma_1^4}, 
\end{eqnarray}
which characterize the skewness of fluctuating fields and  their
derivatives. The quantity $\sigma_i$ characterizes the variance of
fluctuating fields and their derivatives, and is given by 
\begin{equation}
\sigma_j^2\equiv \int_0^\infty\frac{k^2dk}{2\pi^2}k^{2j}P(k)W^2(kR),
\end{equation}
for $d=3$, and
\begin{equation}
\sigma_j^2\equiv \frac1{4\pi}\sum_l(2l+1)\left[l(l+1)\right]^j C_l W^2_l,
\end{equation}
for $d=2$. In both cases $W$ represents a smoothing kernel, or a window
function, which will be given by a product of the experimental beam
transfer function, pixelization window function, and an extra Gaussian
smoothing. The power spectra, $P(k)$ for $d=3$ and $C_l$ for $d=2$, are
defined as 
\begin{eqnarray}
 \langle \tilde{f}_{\mathbf k}\tilde{f}^*_{{\mathbf k}'}\rangle
&=&(2\pi)^3P(k)\delta_D({\mathbf k}-{\mathbf k}'),\\
  \langle a_{lm}a_{l'm'}^*\rangle
&=&C_l\delta_{ll'}\delta_{mm'},
\end{eqnarray}
where $\delta_D({\mathbf k})$ is the Dirac delta function, and the
Fourier and harmonic coefficients are given by
\begin{eqnarray}
 f({\mathbf x})=\int\frac{d^3{\mathbf k}}{(2\pi)^3} \tilde{f}_{\mathbf k}
e^{i{\mathbf k}\cdot{\mathbf x}},\\ 
 f({\mathbf \Omega})=\sum_{lm}a_{lm}Y_{lm}({\mathbf \Omega}),
\end{eqnarray}
for $d=3$ and 2, respectively. Finally, the most relevant Hermite
polynomials are given by
\begin{eqnarray}
H_{-1}(\nu) &=& \sqrt{\frac{\pi}2}e^{\nu^2/2}{\rm
 erfc}\left(\frac{\nu}{\sqrt{2}}\right),\\ 
H_0&=&1,\\
H_1(\nu)&=&\nu,\\
H_2(\nu)&=&\nu^2-1,\\
H_3(\nu)&=&\nu^3-3\nu,\\
H_4(\nu)&=&\nu^4-6\nu^2+3,\\
H_5(\nu)&=&\nu^5-10\nu^3+15\nu.
\end{eqnarray}

\section{Application I: Cosmic Microwave Background}
\label{sec:mf_cmb}
\subsection{Analytical Formula for Minkowski Functionals of CMB}
For the cosmic microwave background, we have $d=2$ and $f=\Delta
T/T$. We define the angular bispectrum as
\begin{equation}
B_{l_1l_2l_3}^{m_1m_2m_3}\equiv
\langle a_{l_1m_1}a_{l_2m_2}a_{l_3m_3}\rangle.
\end{equation}
Then, by expanding skewness parameters into spherical harmonics, we
obtain
\begin{eqnarray}
\label{eq:scmb0}
S^{(0)}
&=&
\frac1{4\pi\sigma_0^4}\sum_{l_im_i}
B_{l_1l_2l_3}^{m_1m_2m_3}{\cal G}_{l_1l_2l_3}^{m_1m_2m_3}
W_{l_1}W_{l_2}W_{l_3},\\
\label{eq:scmb1}
S^{(1)}
&=&
\nonumber
\frac3{16\pi\sigma_0^2\sigma_1^2}\sum_{l_im_i}
\frac{l_1(l_1+1)+l_2(l_2+1)+l_3(l_3+1)}3 \\
& &\times
B_{l_1l_2l_3}^{m_1m_2m_3}{\cal G}_{l_1l_2l_3}^{m_1m_2m_3}
W_{l_1}W_{l_2}W_{l_3}, \\
S^{(2)}
&=&
\nonumber
\frac3{8\pi\sigma_1^4}\sum_{l_im_i}
\left\{
\frac{\left[l_1(l_1+1)+l_2(l_2+1)-l_3(l_3+1)\right]}3
\right. \\
& &
\nonumber
\left.\times l_3(l_3+1)+({\rm cyc.})\right\}
B_{l_1l_2l_3}^{m_1m_2m_3}{\cal G}_{l_1l_2l_3}^{m_1m_2m_3}\\
\label{eq:scmb2}
& &\times W_{l_1}W_{l_2}W_{l_3},
\end{eqnarray}
where (cyc.) means the addition of terms with the same cyclic order of
the subscripts as the previous term, $W_l$ is a smoothing kernel in
$l$ space and ${\cal G}_{l_1l_2l_3}^{m_1m_2m_3}$ is the Gaunt
integral,
\begin{equation}
{\cal G}_{l_1l_2l_3}^{m_1m_2m_3}
\equiv
\int d\hat{\mathbf n}~
Y_{l_1m_1}(\hat{\mathbf n})Y_{l_2m_2}(\hat{\mathbf
n})Y_{l_3m_3}(\hat{\mathbf n}). 
\end{equation}
Note that we have used the following properties of $Y_{lm}(\hat{\mathbf n})$:
\begin{eqnarray}
&\nabla^2 & Y_{lm}(\hat{\mathbf n}) = -l(l+1)Y_{lm}(\hat{\mathbf n}), \\
\nonumber
\int & d\hat{\mathbf n} & \left[\nabla Y_{l_1m_1}(\hat{\mathbf
		       n})\right]\cdot\left[\nabla
		       Y_{l_2m_2}(\hat{\mathbf
		       n})\right]Y_{l_3m_3}(\hat{\mathbf n}) \\
&=&
\frac{l_1(l_1+1)+l_2(l_2+1)-l_3(l_3+1)}2{\cal G}_{l_1l_2l_3}^{m_1m_2m_3}.
\end{eqnarray}
The summation over $m_i$ can be done by writing
\begin{equation}
B_{l_1l_2l_3}^{m_1m_2m_3}={\cal G}_{l_1l_2l_3}^{m_1m_2m_3}b_{l_1l_2l_3},
\end{equation}
where $b_{l_1l_2l_3}$ is the reduced bispectrum that depends on specific
non-Gaussian models \citep{KS2001}. Using the reduced bispectrum, we
finally obtain the analytical formula for MFs of the CMB: 
\begin{eqnarray}
S^{(0)}
&=&
\frac3{2\pi\sigma_0^4}\sum_{2\le l_1\le l_2\le l_3}
I^2_{l_1l_2l_3}b_{l_1l_2l_3}W_{l_1}W_{l_2}W_{l_3}, \label{eq:s0_cmb}\\
S^{(1)}
&=&
\nonumber
\frac3{8\pi\sigma_0^2\sigma_1^2}\sum_{2\le l_1\le l_2\le l_3}
\left[l_1(l_1+1)+l_2(l_2+1)+l_3(l_3+1)\right] \\
& &\times
I^2_{l_1l_2l_3}b_{l_1l_2l_3}W_{l_1}W_{l_2}W_{l_3}, \label{eq:s1_cmb}\\
S^{(2)}
&=&
\nonumber
\frac3{4\pi\sigma_1^4}\sum_{2\le l_1\le l_2\le l_3}
\left\{\left[l_1(l_1+1)+l_2(l_2+1)-l_3(l_3+1)\right]\right. \\
& &
\left.\times l_3(l_3+1)+({\rm cyc.})\right\}
I^2_{l_1l_2l_3}b_{l_1l_2l_3}W_{l_1}W_{l_2}W_{l_3}, \label{eq:s2_cmb}
\end{eqnarray}
where
\begin{equation}
I_{l_1l_2l_3}\equiv
\sqrt{\frac{(2l_1+1)(2l_2+1)(2l_3+1)}{4\pi}}
\left(
\begin{array}{ccc}l_1&l_2&l_3\\0&0&0\end{array}
\right),
\end{equation}
and we have used
\begin{equation}
\sum_{m_1m_2m_3}
\left({\cal G}_{l_1l_2l_3}^{m_1m_2m_3}\right)^2
=
I_{l_1l_2l_3}^2.
\end{equation}

When $f_{\rm NL}$ is a constant, the form of $b_{l_1l_2l_3}$ is given by
\citep{KS2001} 
\begin{eqnarray}
\nonumber
b_{l_1l_2l_3}&=&
2\int^\infty_0r^2dr[b^L_{l_1}(r)b^L_{l_2}(r)b^{NL}_{l_3}(r)
+b^L_{l_1}(r)b^{NL}_{l_2}(r)b^L_{l_3}(r) \\
& &+b^{NL}_{l_1}(r)b^L_{l_2}(r)b^L_{l_3}(r)],
\end{eqnarray}
where 
\begin{eqnarray}
b^L_l(r)\equiv \frac{2}{\pi}\int^\infty_0k^2dkP_\phi(k)g_{Tl}(k)j_l(kr), \\
b^{NL}_l(r)\equiv \frac{2}{\pi}\int^\infty_0k^2dkf_{NL}g_{Tl}(k)j_l(kr),
\end{eqnarray}
and $P_\phi(k)\propto k^{n_s-4}$ is the primordial power spectrum of
$\phi$. The amplitude of $P_\phi(k)$ is fixed by the first peak
amplitude of the temperature power spectrum,
$l(l+1)C_l/(2\pi)=(74.7~\mu{\rm K})^2$ at $l=220$ \citep{WMAPbeam}, and
the temperature power spectrum is given by 
\begin{equation}
 C_l = \frac{2}{\pi}\int^\infty_0k^2dkP_\phi(k)g_{Tl}^2(k).
\end{equation}
We calculate the full radiation transfer function, $g_{Tl}(k)$, using
the publicly-available CMBFAST code \citep{SZ1996}. Note that our
formalism is completely generic. One can easily generalize our results
to non-Gaussian models with a momentum-dependent $f_{\rm NL}$ by using
an appropriate form of $b_{l_1l_2l_3}$ given in \citet{Liguori2006}. 
Our results suggest that the MFs do not contain information beyond the
bispectrum when non-Gaussianity is weak. The MFs of CMB basically
measure the weighted sum of the CMB angular bispectrum.  

\begin{figure*}[t]
\begin{center}
\includegraphics[width=8cm]{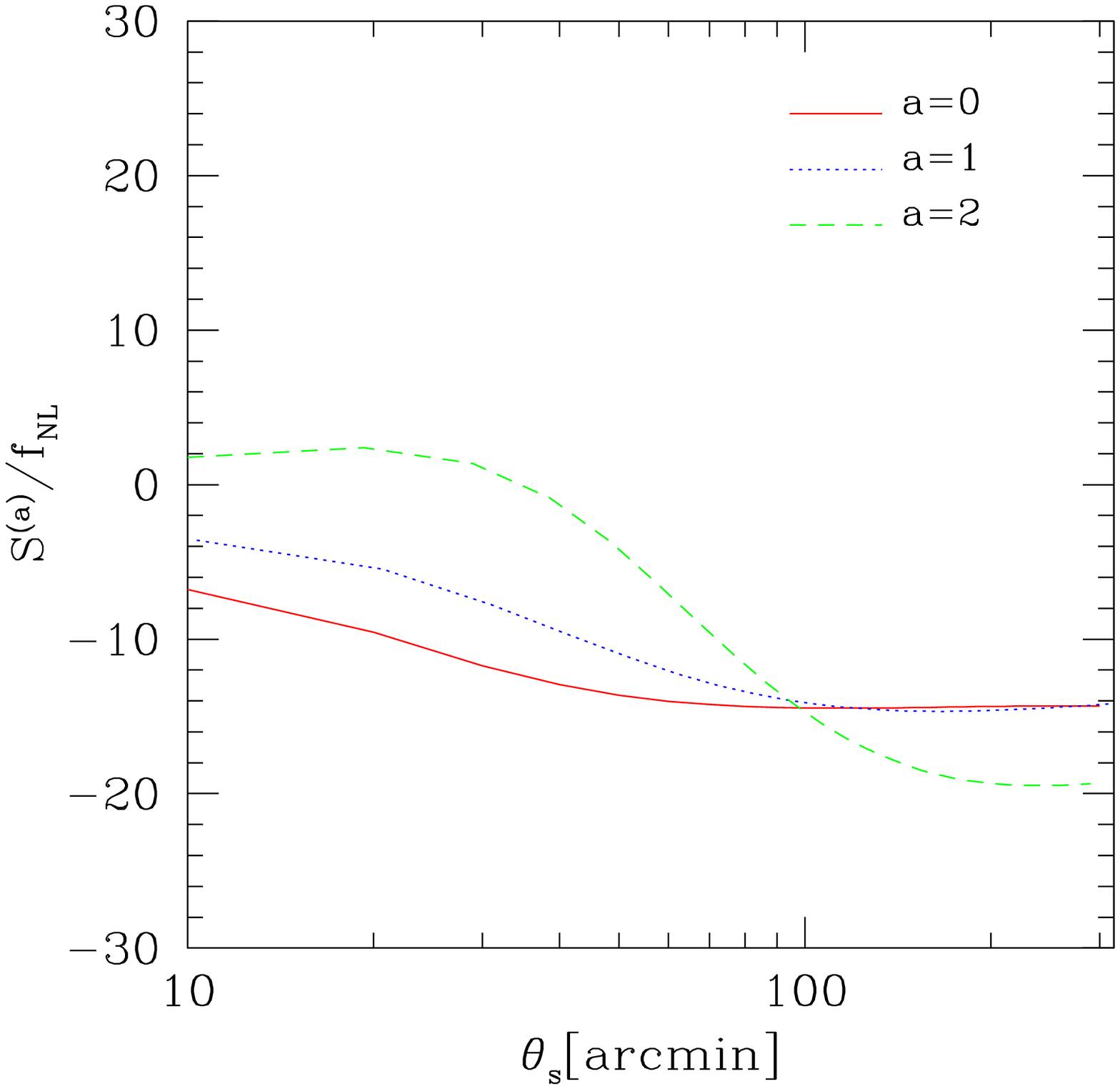}
\includegraphics[width=8.3cm]{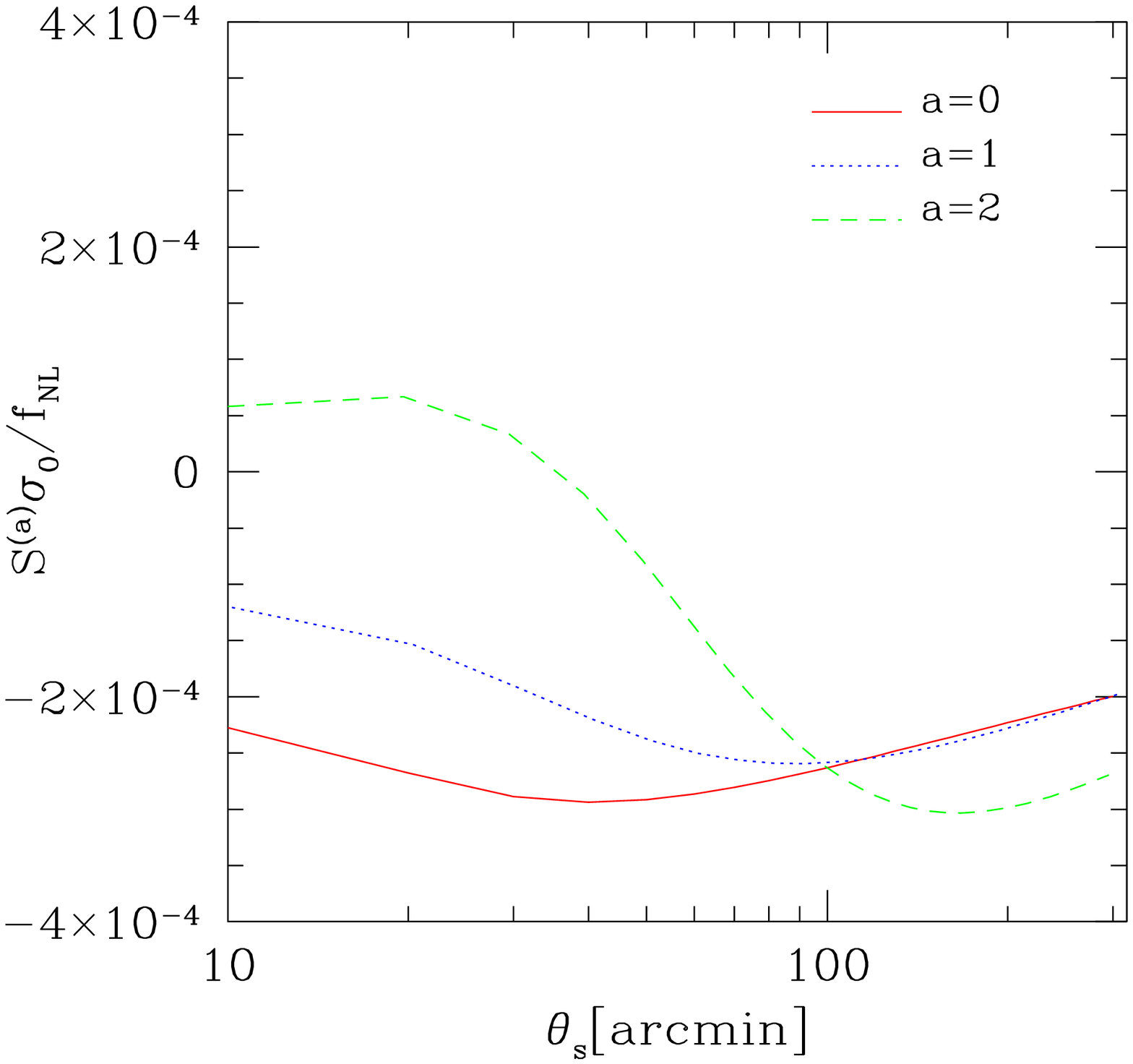}
\caption{%
  ({\it Left}) Skewness parameters, $S^{(a)}$, for $a=0$ (solid), 1
 (dotted), and 2 (dashed).
  ({\it Right}) Skewness parameters multiplied by variance,
 $S^{(a)}\sigma_0$, for $a=0$ (solid), 1  (dotted), and 2 (dashed). Both
 have been divided by $f_{\rm NL}$, and are  plotted as a function of a
 Gaussian smoothing width, $\theta_s$. Note  that $S^{(2)}$ changes its
 sign at $\theta_s\sim 40$~arcmin. 
}%
\label{fig:cmbskew}
\end{center}
\end{figure*}

In Figure \ref{fig:cmbskew}, we plot the skewness parameters,
$S^{(a)}$ (Eqs.~[\ref{eq:s0_cmb}--\ref{eq:s2_cmb}]), 
and $S^{(a)}$ multiplied by
$\sigma_0$, for a pure signal of CMB anisotropy (without noise) 
as a function of a Gaussian smoothing width, $\theta_s$, which determines
a Gaussian smoothing kernel, $W_l=\exp[-l(l+1)\theta_s^2/2]$.
The perturbative expansion of MFs works only when
$S^{(a)}\sigma_0$ is much smaller than unity (see
Eq.~[\ref{eq:MFs_perturb}]).  We find that the perturbative expansion 
is valid for $f_{\rm NL}\ll 3300$ from the results plotted in
the right panel of Figure~\ref{fig:cmbskew} that show
$|S^{(a)}|\sigma_0\lesssim 3\times 10^{-4}~f_{\rm NL}$.

\begin{figure*}
\begin{center}
\includegraphics[width=13cm]{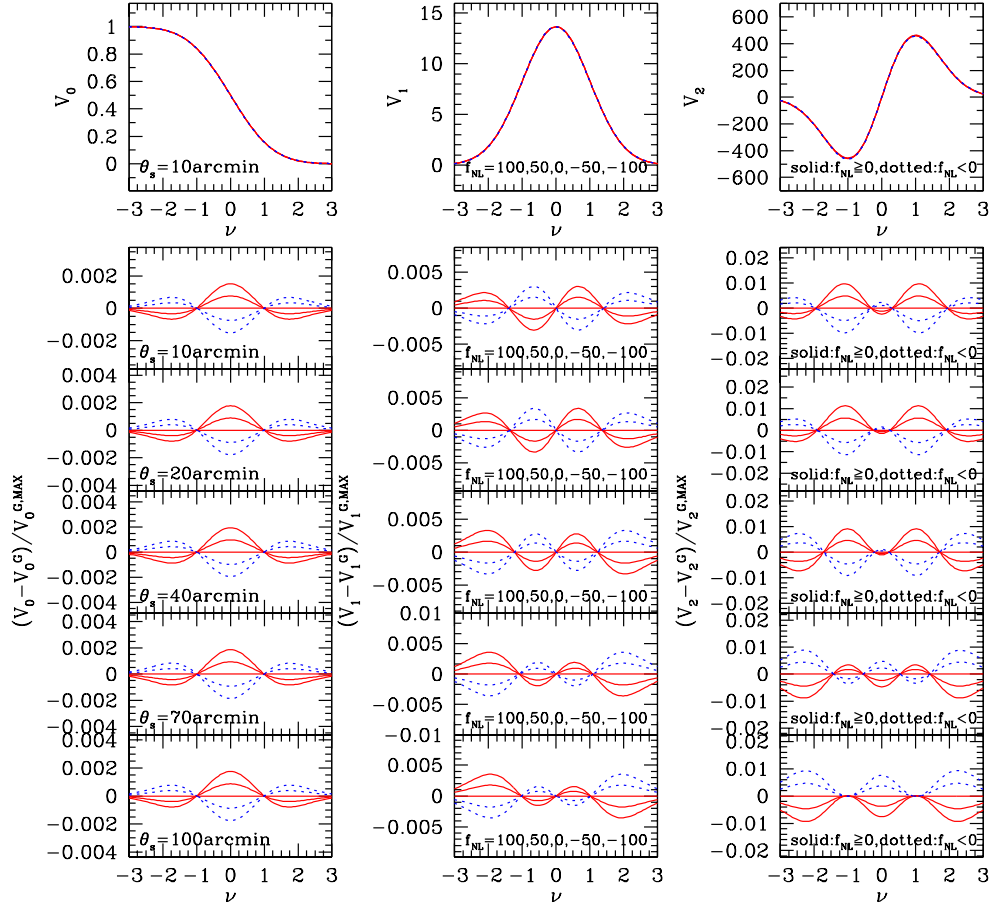}
\caption{%
  Analytical predictions for the Minkowski Functionals of CMB
 temperature anisotropy with primordial non-Gaussianity characterized by
 $f_{\rm NL}=-100$, $-50$ (dotted), 0, 50, and 100 (solid). Each MF,
 $V_{k}$ ($k=0$, 1, and 2), is plotted in the top panels. The other panels
 show the difference between non-Gaussian and Gaussian MFs, $V_k^G$, 
 divided by the maximum amplitude of $V_k^G$. From the
 top to bottom, $\theta_s=10$, 20, 40, 70, and 100~arcmin are shown. 
 }%
\label{fig:cmbmf}
\end{center}
\end{figure*}

This Figure also shows how MFs may be as powerful as the angular
bispectrum in measuring $f_{\rm NL}$.  \citet{KS2001} have shown that
sensitivity of the first skewness parameter, $S^{(0)}$, to $f_{\rm
NL}$ is much worse than that of the angular bispectrum, as acoustic
oscillations in $l$ space smear out non-Gaussian signals in the
skewness, which is the weighted sum of the angular bispectrum over
$l$. (The angular bispectrum is negative in the Sachs-Wolfe regime at
low $l$, and oscillates about zero by changing its sign at higher
$l$.) The MFs are sensitive to not only $S^{(0)}$, but also $S^{(1)}$
and $S^{(2)}$. The weight of the sum over multipoles differs among
three skewness parameters: $S^{(2)}$ has the largest weight at high
$l$, $S^{(0)}$ has the largest weight at low $l$, and $S^{(1)}$ is
somewhere in between (Eqs.~[\ref{eq:s0_cmb}--\ref{eq:s2_cmb}]). In
particular, because $S^{(2)}$ picks up the highest multipoles
efficiently, $S^{(2)}$ changes its sign depending on $\theta_s$.
$S^{(2)}$ is negative on very large angular scales.  As $\theta_s$
decreases (as the small scale information is included), $S^{(2)}$
increases, and eventually changes its sign to positive values near
the scale of the first acoustic peak, $\theta_s\sim 40$~arcmin, where the
bispectrum has the largest amplitude, while the other two skewness
parameters do not change their signs. Therefore, $S^{(2)}$ keeps
information about the acoustic oscillations. This property is crucial
for obtaining a better signal-to-noise ratio for primordial
non-Gaussianity in the CMB.  

Figure \ref{fig:cmbmf} shows the predicted MFs of CMB temperature
anisotropy, $V_0$, $V_1$ and $V_2$, as a function of $f_{\rm NL}$.  The
MFs for $f_{\rm NL}=100, 50$ and $0$ are plotted in the solid lines,
while the MFs for $f_{\rm NL}=-50$ and $-100$ are plotted in the dotted
lines. The lower panels show the difference between the Gaussian and
non-Gaussian MFs divided by the maximum amplitude of Gaussian MFs. 
Primordial non-Gaussianity with $f_{\rm NL}=100$ changes $V_0$, $V_1$
and $V_2$ by 0.2\%, 0.4\%, and 1\%, respectively, relative to the
maximum amplitude of the corresponding Gaussian MFs. While $V_0$
(area) has little dependence on $\theta_s$, $V_2$ (Euler characteristic)
depends on $\theta_s$ strongly, which is mainly due to the sign change
of $S^{(2)}$ at $\theta_s\sim 40$~arcmin. 

In Appendix~\ref{app:cmbsim} we show that these analytical predictions  
agree with non-Gaussian simulations in the Sachs--Wolfe limit very well.  
The comparison with the full simulations that include the full radiation
transfer function will be reported elsewhere.

\subsection{Measuring $f_{\rm NL}$ from Minkowski Functionals of CMB}
\label{subsec:fnl_cmb}
The MFs at different threshold values, $\nu$, are correlated,
and different MFs are also correlated, i.e.,
\begin{equation}
\Sigma_{kk'}^{(d)}(\nu,\nu') \equiv
\langle V_k^{(d)}(\nu)V_{k'}^{(d)}(\nu')\rangle 
\neq \delta_{kk'}\delta(\nu-\nu').
\end{equation}
where $\delta$ denotes the Kronecker delta. Therefore, it is important
to use the full covariance matrix, $\Sigma$, in the data analysis. We
obtain the full covariance matrix from Monte Carlo simulations of {\it
Gaussian} temperature anisotropy. Because non-Gaussianity is weak, the
covariance matrix estimated from Gaussian simulations is a good
approximation of the exact one. In Appendix~\ref{app:comp_cmb} we
describe our methods for computing MFs from the pixelized CMB maps. In
Appendix~\ref{app:sim_wmap} and \ref{app:sim_planck} we describe our
methods for simulating sky maps of CMB temperature anisotropy with
noise and instrumental characteristics of the WMAP and Planck 
experiment respectively.

We use the Fisher information matrix formalism to estimate the 
projected errors on $f_{\rm NL}$ from given measurement errors on MFs.
The Fisher matrix, $F_{ij}$, is written in terms of the inverse of the
covariance matrix, $\Sigma^{-1}$, as
\begin{equation}
\label{eq:covmatrix0}
  F^{(d)}_{ij}=\sum_{kk'}\int d\nu\int d\nu'~ \frac{\partial
  V^{(d)}_k(\nu)}{\partial
  p_i}(\Sigma^{-1})^{(d)}_{kk'}(\nu,\nu')\frac{\partial
  V^{(d)}_{k'}(\nu')}{\partial p_j}, 
\end{equation}
where $p_i$ is the $i$-th parameter.  For CMB, we consider only one
parameter, $p_1=f_{NL}$, whereas for LSS we also include galaxy
biasing parameters (see \S~\ref{sec:mf_lss}).  The projected
1-$\sigma$ error on $f_{NL}$ is given by the square root of
$(F^{-1})_{11}$. While equation~(\ref{eq:covmatrix0}) may be evaluated
for a given smoothing scale, $\theta_s$, one will eventually need to
combine all combinations of MFs at different $\theta_s$ to obtain the
best constraint on $f_{\rm NL}$ from given data. The MFs at different
$\theta_s$ are also correlated. We therefore calculate the full
covariance matrix of MFs that consists of $N_\nu \times N_{\rm MF}\times
N_{\rm s}$ elements, where $N_\nu=25$ is the number of bins for $\nu$
per each MF in the range of $\nu$ from $-3$ to $3$, $N_{\rm MF}=3$ and
$4$ is the number of MFs for $d=2$ and $3$, respectively (i.e., $N_{\rm
MF}=3$ for CMB and $4$ for LSS), and $N_{\rm s}$ is the number of
smoothing scales used in the analysis. The Fisher matrix may be written as
\begin{equation}
\label{eq:covmatrix}
  F^{(d)}_{ij}=\sum_{\alpha\alpha'}\frac{\partial
  V^{(d)}_\alpha}{\partial
  p_i}(\Sigma^{-1})^{(d)}_{\alpha\alpha'}\frac{\partial
  V^{(d)}_{\alpha'}}{\partial p_j}, 
\end{equation}
where $\alpha$ is a single index denoting $k$, $\nu$, and $\theta_s$.

\begin{table*}[t]
\caption{%
  Projected 1-$\sigma$ errors on $f_{\rm NL}$ from each of the Minkowski
 Functionals of CMB temperature anisotropy ($V_0$: area, $V_1$: total
 circumference, $V_2$: Euler characteristic) as well as from the
 combined analysis of all the Minkowski Functionals,  for various
 combinations of smoothing scales, $\theta_s$. We use noise and beam
 properties of WMAP 1-year and 8-year observations with the $Kp0$
 mask. The last block of Table shows the 1-$\sigma$ errors from the
 ``ultimate WMAP'', which uses all sky with zero noise. (The beam
 smearing is still included.) 
}%
\begin{center}
\begin{tabular}{cclclcl}
  \hline\hline
 & \multicolumn{2}{c}{WMAP 1-year} & \multicolumn{2}{c}{WMAP 8-year}
 & \multicolumn{2}{c}{no noise and no sky cut}  \\ 
\cline{2-7}
 \raisebox{1.5ex}[0pt]{$\theta_s$[arcmin]} & All & ($V_0,V_1,V_2$) & All & ($V_0,V_1,V_2$) & All & ($V_0,V_1,V_2$)\\
  \hline
$       100$ & $ 271$ & $( 671, 297, 494)$ & $ 269$ & $( 668, 296, 494)$ & $ 135$ & $( 461, 146, 182)$ \\
$        70$ & $ 153$ & $( 456, 168, 272)$ & $ 153$ & $( 456, 168, 271)$ & $  94$ & $( 363, 100, 137)$ \\
$        40$ & $  86$ & $( 283,  96, 148)$ & $  84$ & $( 289,  95, 144)$ & $  59$ & $( 236,  67,  94)$ \\
$        20$ & $  54$ & $( 176,  67,  82)$ & $  51$ & $( 183,  63,  79)$ & $  41$ & $( 160,  52,  62)$ \\
$        10$ & $  73$ & $( 139,  92,  91)$ & $  41$ & $( 134,  54,  61)$ & $  35$ & $( 120,  47,  51)$ \\
$         5$ &    --- &                    & $  75$ & $( 125,  94,  87)$ & $  30$ & $( 106,  42,  43)$ \\
$  10,20~\&~40$ & $  46$ & $(  92,  62,  67)$ & $  36$ & $(  82,  51,  56)$ & $  30$ & $(  49,  43,  45)$ \\
$   5,10~\&~20$ &    --- &                    & $  36$ & $(  71,  52,  53)$ & $  22$ & $(  46,  32,  37)$ \\
\hline
\end{tabular} 
\end{center}
\label{tab:fnl_cmb_wmap}
\end{table*}
\begin{table*}[t]
\caption{%
Same as Table \ref{tab:fnl_cmb_wmap} but for Planck data. 
}%
\begin{center}
\begin{tabular}{cclcl}
  \hline\hline
 & \multicolumn{2}{c}{Planck} & \multicolumn{2}{c}{no noise and no sky cut}  \\ 
\cline{2-5}
 \raisebox{1.5ex}[0pt]{$\theta_s$[arcmin]} & All & ($V_0,V_1,V_2$) & All & ($V_0,V_1,V_2$)\\
  \hline
$       100$ & $ 271$ & $( 665, 299, 502)$ & $ 136 $ & $( 462, 147, 184)$ \\
$        70$ & $ 157$ & $( 463, 171, 278)$ & $  95 $ & $( 354, 102, 141)$ \\
$        40$ & $  77$ & $( 281,  86, 128)$ & $  54 $ & $( 235,  61,  84)$ \\
$        20$ & $  46$ & $( 153,  59,  69)$ & $  37 $ & $( 140,  49,  55)$ \\
$        10$ & $  32$ & $( 115,  45,  45)$ & $  28 $ & $( 102,  40,  39)$ \\
$         5$ & $  24$ & $( 97,  31,  36)$ & $  21 $ & $(  87,  27,  30)$ \\
$  5,10~\&~20$ & $  19$ & $(  46,  26,  31)$ & $  15 $ & $(  38,  22,  26)$ \\
\hline
\end{tabular} 
\end{center}
\label{tab:fnl_cmb_planck}
\end{table*}

Let us comment on the effect of noise on MFs. The instrumental noise
increases $\sigma_j^2$ by adding extra power at small scales. On the
other hand, the signal part of the angular bispectrum is unaffected by
noise because noise is Gaussian. As a result, the instrumental noise
always reduces the skewness parameters, as the skewness parameters
contain $\sigma_j$ in their denominator. Therefore, the MFs would
approach Gaussian predictions in the noise-dominated limit, as
expected. We compute the increase in $\sigma_0$ and $\sigma_1$ due to
noise from Monte Carlo simulations, and then rescale $S^{(a)}$ and
$\partial V_\alpha/\partial f_{\rm NL}$ for a given smoothing scale,
$\theta_s$. The window function, $W_l$, includes the beam smearing
effect, pixel window function, and a Gaussian smoothing.

Using the method described above and in Appendix~\ref{app:sim}, we
estimate the projected 1-$\sigma$ error on $f_{NL}$ expected from
WMAP's 1-year and 8-year observations. We also consider an
ideal WMAP experiment without noise or sky cut (but the
beam smearing is still included).  We consider six
different smoothing scales, $\theta_s=5, 10, 20, 40, 70$, and
100~arcmin. The results from various combinations of $\theta_s$ are
summarized in Table~\ref{tab:fnl_cmb_wmap} for WMAP observations. 

For the WMAP 1-year data, the error on $f_{\rm NL}$ is the smallest for
$\theta_s=20$ arc-minutes. At smaller angular scales, say
$\theta_s=10$ arc-minutes, the noise dominates more and thus the error
on $f_{\rm NL}$ increases at $\theta_s\lesssim 20$~arc-minutes. 
For the WMAP 8-year
data, on the other hand, a better signal-to-noise ratio at smaller
angular scales enables us to constrain MFs at $\theta_s=10$~arc-minutes.
As the beam size of WMAP in W band is about 
$\theta_s=10$~arc-minutes\footnote{Note that $\theta_s$ is a Gaussian width, which is 1/2.35 times the
full width at half maximum.},
one cannot constrain MFs at the angular scales smaller than this.
When all the smoothing scales are combined,
the projected 1-$\sigma$ error on $f_{\rm NL}$ reaches $\sim 40$ for
the WMAP data, which is in a rough agreement with the result reported
in \citet{Komatsu2003} and \citet{Spergel2006}. The best constraint
that can be obtained from the WMAP data, in the limit of zero noise
and full sky coverage, is $f_{\rm NL}\sim 22$.  

We also estimate the Planck constraint on $f_{\rm NL}$ listed at Table
\ref{tab:fnl_cmb_planck}. As Planck's beam and noise are $\sim 4$ and
10 times as small as WMAP's, respectively, one can constrain MFs even
at $\theta_s=5$~arc-minutes. Planck should be sensitive to $|f_{\rm
NL}|\sim 20$.

\section{Application II: Large-scale Structure}\label{sec:mf_lss}
\subsection{Analytical Formula for Minkowski Functionals of LSS}
For the large-scale structure, we have $d=3$ and $f=\delta_{\rm
g}(\mathbf{x}, z)$, where $\delta_{\rm g}$ is the density contrast of
galaxies. Statistical isotropy of the universe gives the following form
of the bispectrum:
\begin{eqnarray}
\nonumber
 \langle\tilde{\delta}_{\rm g}(z)({\mathbf k_1})\tilde{\delta}_{\rm
  g}(z)({\mathbf k_2})\tilde{\delta}_{\rm g}(z)({\mathbf k_3})\rangle
  &\equiv & (2\pi)^3\delta_D({\mathbf k_1}+{\mathbf k_2}+{\mathbf k_3}) \\
  & &\times B_{\rm g}(k_1,k_2,k_3,z). 
\end{eqnarray}
We obtain the skewness parameters by integrating 
$B_{\rm g}(k_1,k_2,k_3,z)$ over 
$k_1$, $k_2$, and $\mu\equiv({\mathbf k_1}\cdot{\mathbf 
k_2})/(k_1k_2)$ with appropriate weights as
\begin{eqnarray}
\label{eq:s0_lss}
\nonumber
S^{(0)}_g(z)&=&\frac{1}{8\pi^4\sigma_{{\rm g},0}^4(z)}
\int^\infty_0 dk_1 \int^\infty_0 dk_2 \int^1_{-1} d\mu \\
& &\times
\nonumber
k_1^2k_2^2 B_{\rm g}(k_1,k_2,k_{12},z) \\
& &\times W(k_1R)W(k_2R)W(k_{12}R),\\
\nonumber
S^{(1)}_g(z)&=&
\frac{1}{16\pi^4\sigma_{{\rm g},0}^2(z)\sigma_{{\rm g},1}^2(z)}
\int^\infty_0 dk_1 \int^\infty_0 dk_2 \int^1_{-1} d\mu \\
& &\times
\nonumber
k_1^2k_2^2(k_1^2+k_2^2+\mu k_1k_2)
B_{\rm g}(k_1,k_2,k_{12},z)\\
\label{eq:s1_lss}
& &\times W(k_1R)W(k_2R)W(k_{12}R),\\
\nonumber
S^{(2)}_g(z)&=&\frac{3}{16\pi^4\sigma_{{\rm g},1}^4(z)}
\int^\infty_0 dk_1 \int^\infty_0 dk_2 \int^1_{-1} d\mu \\
& &\times
\nonumber
k_1^4k_2^4 (1-\mu^2)B_{\rm g}(k_1,k_2,k_{12},z)\\
\label{eq:s2_lss}
& &\times W(k_1R)W(k_2R)W(k_{12}R),
\end{eqnarray}
where $k_{12}\equiv|{\mathbf k_1}+{\mathbf k_2}|=(k_1^2 +k_2^2+2\mu
k_1k_2)^{1/2}$. 

Unlike for the CMB, where we needed to consider only the
effect of primordial non-Gaussianity, there are three sources of
non-Gaussianity in $B_{\rm g}(k_1,k_2,k_3,z)$: primordial
non-Gaussianity, non-linearity in the gravitational evolution, and
non-linearity in the galaxy bias. 
In Appendix~\ref{app:Bg} we show that $B_{\rm g}$ is given by
\begin{eqnarray}
\nonumber
 B_{\rm g}(k_1,k_2,k_3,z)
&=&
b_1^3(z)\left[B_{\rm prim}(k_1,k_2,k_3,z)\right. \\
& &
\nonumber
\left.+B_{\rm grav}(k_1,k_2,k_3,z)\right] +b_1^2(z)b_2(z)\\
& &\times
\left[P_{\rm m}(k_1,z)P_{\rm m}(k_2,z)+(\mbox{cyc.})\right],
\end{eqnarray}
where $P_{\rm m}(k,z)$ is the linear matter power spectrum, $b_1(z)$ and
$b_2(z)$ are the linear and non-linear galaxy bias parameters,
respectively (see Eq.~[\ref{eq:fg93}] for the precise definition), and
$B_{\rm prim}$ and $B_{\rm grav}$ represent the contributions from
primordial non-Gaussianity and non-linearity in gravitational
clustering, respectively: 
\begin{eqnarray}
\nonumber
 B_{\rm prim}(k_1,k_2,k_3,z)
&\equiv&
 \frac{2f_{\rm NL}}{D(z)}
\left[\frac{P_{\rm m}(k_1,z)P_{\rm m}(k_2,z)M(k_3)}{M(k_1)M(k_2)}
\right. \\
& & \left.+ (\mbox{cyc.})\right],\\
B_{\rm grav}(k_1,k_2,k_3,z) 
&\equiv&
\nonumber
 2\left[F_2({\mathbf k}_1,{\mathbf k}_2)P_{\rm m}(k_1,z) 
P_{\rm m}(k_2,z)\right. \\
& &
\left.+(\mbox{cyc.})\right],
\end{eqnarray}
where $D(z)$ is the growth rate of linear density fluctuations
normalized such that $D(z)\rightarrow 1/(1+z)$ during the matter era,
and $M(k)$ and $F_2({\mathbf k}_1,{\mathbf k}_2)$ are
given by equation~(\ref{eq:Mk}) and (\ref{eq:f2}), respectively.
These equations suggest that $f_{\rm NL}$ and the galaxy
bias parameters must be determined simultaneously from the LSS
data. Moreover, even if the galaxy bias is perfectly linear,
$b_2\equiv 0$, the primordial signal might be swamped by
non-Gaussianity due to non-linear gravitational clustering,
$B_{\rm grav}$.

\begin{figure*}[ht]
\begin{center}
\includegraphics[width=8cm]{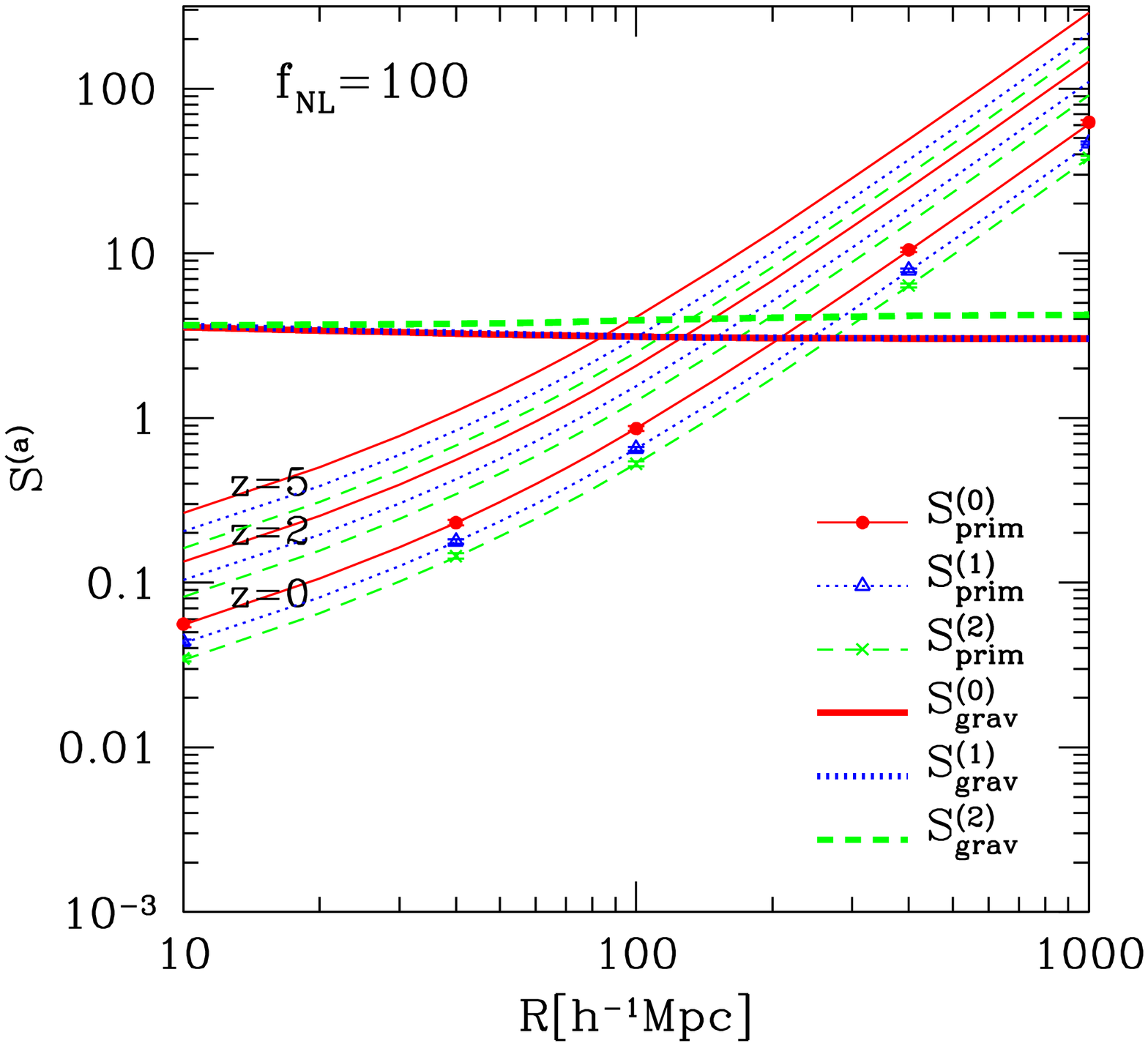}
\includegraphics[width=8cm]{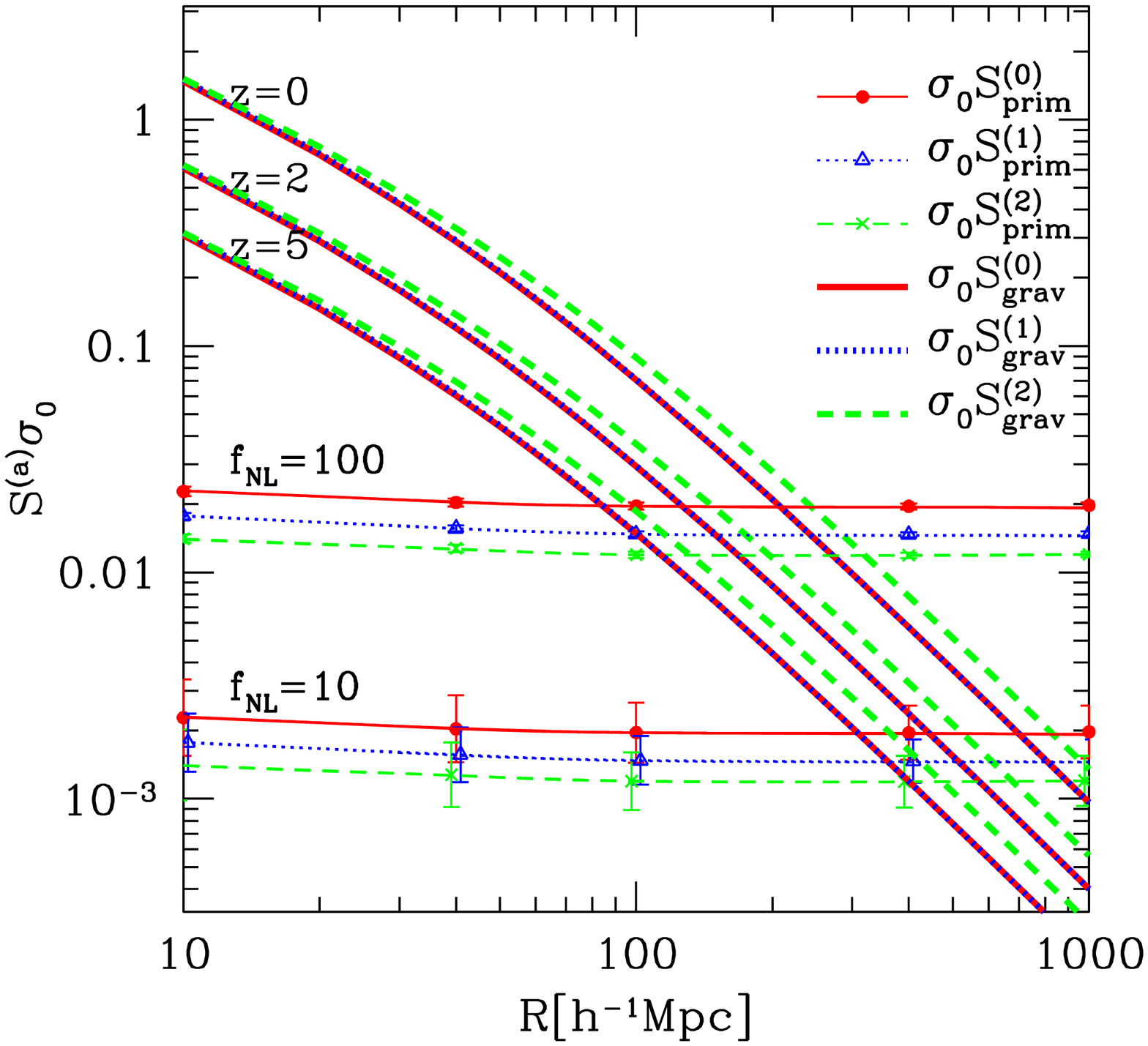}
\caption{%
  ({\it Left}) Skewness parameters, $S^{(a)}(z)$, for $a=0$ (solid), 1
 (dotted), and 2 (dashed), at $z=0$, 2, 5. 
  ({\it Right}) Skewness parameters multiplied variance,
 $S^{(a)}(z)\sigma_0$, for $a=0$ (solid), 1 (dotted), and 2 (dashed), at
 $z=0$, 2, 5. In both figures, the thick lines show the skewness
 parameters from non-linear gravitational clustering, $S_{\rm
 grav}^{(a)}$ (Eq.~[\ref{eq:sgrav0}--\ref{eq:sgrav2}]), while the thin
 lines show the primordial skewness parameters, $S_{\rm prim}^{(a)}$,
 with $f_{\rm NL}=100$ (and $10$ in the right panel). The symbols show
 the skewness parameters measured from numerical simulations of
 non-Gaussian matter density fluctuations. Note that $S_{\rm
 grav}^{(a)}$ and $S_{\rm prim}^{(a)}\sigma_0$  are independent of $z$.  
 }%
\label{fig:skew_lss}
\end{center}
\end{figure*}

In order to investigate how important the effect of $B_{\rm grav}$ is,
 let us define the skewness parameters that are contributed solely by
 $B_{\rm prim}$ or $B_{\rm grav}$. Substituting $B_{\rm grav}$ and
 $\sigma_{{\rm m},j}(z)$ for  $B_{\rm g}$ and $\sigma_{{\rm g},j}$,
 respectively, in equation~(\ref{eq:s0_lss}), (\ref{eq:s1_lss}), and
 (\ref{eq:s2_lss}), we obtain $S^{(a)}_{\rm grav}$ ($a=0$, 1, and 2)
 given by \citep{Matsubara2003} 
\begin{eqnarray}
\nonumber
S^{(0)}_{\rm grav}(z)&=&\frac{3}{28\pi^4\sigma_{{\rm m},0}^4(z)}
\left[5I_{220}(z)
+7I_{131}(z)+2I_{222}(z)\right], \\
\label{eq:sgrav0}
& & \\
\nonumber
S^{(1)}_{\rm grav}(z)&=&\frac{3}{56\pi^4\sigma_{{\rm m},0}^2(z)
\sigma_{{\rm m},1}^2(z)}
\left[10I_{240}(z)+12I_{331}(z) \right. \\
\label{eq:sgrav1}
& &
\left.+7I_{151}(z)+11I_{242}(z)+2I_{333}(z)
\right], \\
\nonumber
S^{(2)}_{\rm grav}(z)&=&\frac{9}{56\pi^4\sigma_{{\rm m},1}^4(z)}
\left[5I_{440}(z)+7I_{351}(z)-3I_{442}(z)\right. \\
\label{eq:sgrav2}
& &
\left. -7I_{353}(z)-2I_{444}(z)\right],
\end{eqnarray}
where 
\begin{eqnarray}
\nonumber
I_{mnr}(z) & \equiv & \int^\infty_0dk_1\int^\infty_0dk_2\int^1_{-1}d\mu
W(k_1R)W(k_2R) \\
& &\times W(k_{12}R)k_1^mk_2^n\mu^rP_{\rm m}(k_1,z)P_{\rm m}(k_2,z).
\end{eqnarray}
Note that $\sigma_{{\rm g},j}=b_1\sigma_{{\rm m},j}$. Similarly, we also
calculate the primordial skewness parameters, $S^{(a)}_{\rm prim}$, by
substituting $B_{\rm prim}$ and $\sigma_{{\rm m},j}(z)$ for  $B_{\rm g}$
and $\sigma_{{\rm g},j}(z)$, respectively, in
equation~(\ref{eq:s0_lss}), (\ref{eq:s1_lss}), and
(\ref{eq:s2_lss}). These skewness parameters are related to the skewness
parameters of the total galaxy bispectrum as 
\begin{equation}
S_{\rm g}^{(a)}=\frac{S^{(a)}_{\rm prim}+S^{(a)}_{\rm
grav}}{b_1}+\frac{3b_2}{b_1^2}.
\end{equation}

In Figure \ref{fig:skew_lss} we compare $S_{\rm prim}^{(a)}(z)$ and
$S_{\rm grav}^{(a)}(z)$ at $z=0$, 2, and 5, as a function of a smoothing
length, $R$, which is in units of $h^{-1}$~Mpc. (The smoothing kernel
$W(kR)$ is set to be a Gaussian filter, $W(kR)=\exp(-k^2R^2/2)$.)
Non-Gaussianity from non-linear gravitational clustering always gives
positively skewed density fluctuations, $S^{(a)}_{\rm grav}>0$. 
Primordial non-Gaussianity with a positive $f_{\rm NL}$ also yields
positively skewed density fluctuations; however, primordial
non-Gaussianity with a negative $f_{\rm NL}$ yields {\it negatively}
skewed density fluctuations, which may be distinguished from
$S^{(a)}_{\rm grav}$ more easily.  As the smoothing scale, $R$,
increases (i.e., density fluctuations become more linear),
non-Gaussianity from non-linear clustering, $S^{(a)}_{\rm
grav}\sigma_{{\rm m},0}$, becomes weaker, while primordial
non-Gaussianity, $S^{(a)}_{\rm prim}\sigma_{{\rm m},0}$, remains nearly
the same. At $z=0$, the primordial contribution exceeds non-linear
gravity only at very large scales, $R>200~h^{-1}$~Mpc for $f_{\rm
NL}=100$, and $R>800~h^{-1}$~Mpc for $f_{\rm NL}=10$. As higher
redshift, on the other hand, non-linearity is much weaker and therefore
the primordial contribution dominates at relatively smaller spatial
scales, $R>120~h^{-1}$~Mpc and $80~h^{-1}$~Mpc at $z=2$ and 5,
respectively, for $f_{\rm NL}=100$. Unlike for the CMB, all the skewness
parameters of galaxies exhibit similar dependence on the smoothing
scales.  The perturbation formula is valid when the amplitude of the
second order correction of MFs is small, $S^{(k)}\sigma_0\ll 1$, that
is, $f_{\rm NL}\ll 5000$.  

\subsection{Measuring $f_{\rm NL}$ from Minkowski Functionals of LSS}
\label{sec:method_lss}
\begin{figure*}
\begin{center}
\includegraphics[width=16cm]{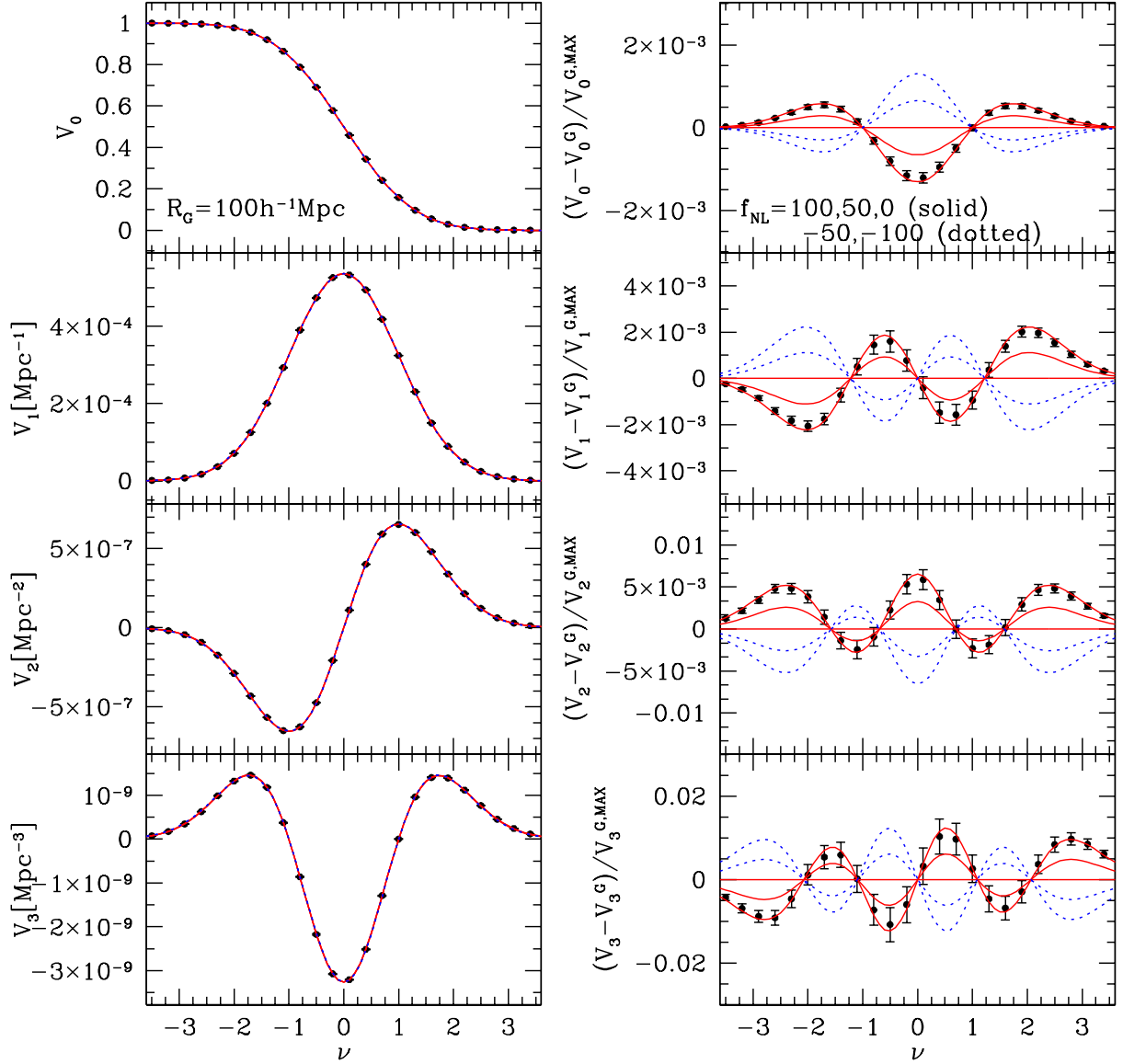}
\caption{%
 Analytical predictions for the Minkowski Functionals of LSS with primordial
 non-Gaussianity characterized by $f_{\rm NL}=100$, 50, 0 (solid),
 $-50$, and $-100$ (dotted), for a Gaussian smoothing length of
 $R=100~h^{-1}$~Mpc. The left panels show the MFs $V_k$ ($k=0$, 1, 2,
 and 3 from top to bottom), while the right panels show the difference
 between non-Gaussian and Gaussian MFs, $V_k^{G}$, divided by the
 maximum amplitude of $V_k^{G}$. The symbols show the average and error
 of the MFs calculated from 2000 realizations of simulated primordial
 non-Gaussian density fluctuations with $f_{\rm NL}=100$.
}%
\label{fig:mf100}
\end{center}
\end{figure*}

Figure \ref{fig:mf100} shows the perturbation predictions for the MFs from
primordial non-Gaussianity with $f_{\rm NL}=100$, 50, 0, $-50$, and
$-100$. For comparison, we also show the MFs computed from numerical
simulations with $f_{\rm NL}=100$. In Appendix~\ref{app:comp_lss} we
describe our methods for computing MFs from the LSS data. In
Appendix~\ref{app:sim_lss} we describe our methods for simulating the
LSS data with primordial non-Gaussianity (but without any effects from
non-linear gravitational clustering or galaxy bias). The error-bars are
estimated from variance among 2000 realizations divided by
$\sqrt{2000}$. The left panels show the MFs, while the right panels show
the difference between the non-Gaussian and Gaussian MFs, divided by the
maximum amplitude of each MF. We find that the analytical perturbation
predictions agree with the numerical simulations very well. 

Let us comment on some subtlety that exists in the comparison between the
perturbation predictions and numerical simulations. The MFs measured
from numerical simulations often deviate from the analytical
predictions, even for Gaussian fluctuations, due to subtle pixelization
effects. The MFs from our Gaussian simulations deviate from the
analytical predictions at the level of $10^{-2}$ when normalized to the
maximum amplitude of each MF. It is important to remove this bias, as
the magnitude of this effect is comparable to or larger than the effect
of primordial non-Gaussianity with $f_{\rm NL}=100$ ($10^{-3}$ for $V_0$
and $10^{-2}$ for $V_3$). Therefore, it is often necessary to
re-calibrate the Gaussian predictions for the pixelization effects by
running a large number of Gaussian realizations. However, we have found
that the {\it difference} between the Gaussian and non-Gaussian MFs
measured from simulations does agree with the perturbation predictions
without any corrections. Therefore, one may use the following procedure
for calculating the correct non-Gaussian MFs:
\begin{itemize}
 \item[(1)] Use the analytical formulae (Eq.~[\ref{eq:MFs_perturb}]) to
	    calculate the difference between the Gaussian and
	    non-Gaussian MFs, 
\begin{equation}
\Delta V_k^{(d)}(\nu, f_{\rm NL})
\equiv V_k^{(d)}(\nu, f_{\rm NL})-V_k^{(d)}(\nu, f_{\rm NL}=0).
\end{equation}
\item[(2)] Run Gaussian simulations. Estimate the average MFs from
	    these Gaussian simulations, $\tilde{V}_k^{(d)}(\nu, f_{\rm
	    NL}=0)$. This 
	    would be slightly different from the analytical formula,
	    $V_k^{(d)}(\nu, f_{\rm NL}=0)$, due to the pixelization and
	    boundary effects. Note that the same simulations may be used
	    to obtain the covariance matrix of MFs.
\item[(3)] Calculate the final non-Gaussian predictions as
\begin{equation}
 \tilde{V}_k^{(d)}(\nu, f_{\rm NL})
= \tilde{V}_k^{(d)}(\nu, f_{\rm NL}=0) + \Delta V_k^{(d)}(\nu, f_{\rm NL}).
\end{equation}
\end{itemize}
We estimate the projected errors on $f_{\rm NL}$ from LSS using the
Fisher information matrix in the same way as CMB (see
\S~\ref{subsec:fnl_cmb}). We compute the covariance matrix of 
MFs from 2000 realizations of Gaussian density fluctuations in a 
1~Gpc$^3$ cubic box. For simplicity, we focus on the large scales and
ignore non-Gaussianity from non-linear gravitational evolution. (The
minimum smoothing scale is $R=60~h^{-1}$~Mpc). We also ignore shot noise.

Table~\ref{tab:fnllimit_lss} shows the projected 1-$\sigma$ errors on
$f_{\rm NL}$ from a galaxy survey covering 1~$h^{-3}~{\rm Gpc}^3$
volume. This volume would correspond to e.g., a galaxy survey covering
300~deg$^2$ on the sky at $5.5<z<6.5$.
Note that these constraints are independent of $z$ when
non-linear gravitational evolution is ignored, as $S_{\rm
prim}^{(a)}\sigma_0$ is independent of $z$. We have assumed a linear
galaxy bias ($b_2=0$) in the second column, while we have marginalized
$b_2/b_1$ in the third column. A more realistic prediction would lie
between these two cases, as we can use the power spectrum and bispectrum
to put some constraints on $b_2$. We find better constraints from
smaller smoothing scales for a given survey volume, simply because we
have more modes on smaller scales. The limits on $f_{\rm NL}$ from a
1~$h^{-3}~{\rm Gpc}^3$ survey are not very promising, $|f_{\rm NL}|\sim
270$ at the 68\% confidence level; thus, one would need the survey
volume as large as 25~$h^{-3}~{\rm Gpc}^3$ to make the LSS constraints
comparable to the WMAP constraints (using the MFs only). This could be
done by a survey covering $\sim 2000$~deg$^2$ at $3.5<z<6.5$. One would
obviously need more volume to make it comparable to the Planck data.  

\begin{table*}[t]
\caption{%
 Constraints on $f_{\rm NL}$ from the MFs of LSS in a galaxy
 survey covering 1~$h^{-3}~{\rm Gpc}^3$ volume. All of the four MFs have
 been combined. The first column shows the smoothing scales that have
 been used. The second column shows the constraints assuming a linear
 bias ($b_2=0$), while the third column shows the constraints with
 $b_2/b_1$ marginalized. Note that these constraints are independent of
 $z$ when non-linear gravitational evolution is ignored, as $S_{\rm
 prim}^{(a)}\sigma_0$ is independent of $z$. 
 }%
\begin{center}
\begin{tabular}{ccc}
  \hline\hline
 $R~h^{-1}$~Mpc & $f_{\rm NL}$ ($b_2/b_1=0$)  & $f_{\rm NL}$ ($b_2/b_1$
 marginalized) \\  
  \hline
  $100, 120~\&~140$  & $540$ & $1200$ \\
   $80, 100~\&~120$ & $350$ & $670$ \\
   $60, 80~\&~100$ & $270$ & $630$  \\ \hline
\end{tabular}
\end{center}
\label{tab:fnllimit_lss}
\end{table*}

\section{Summary and Conclusions}\label{sec:summary}
We have derived analytical formulae of the MFs for CMB and LSS using a
perturbation approach. The analytical formula is useful for studying the
behavior of MFs and estimating the observational constraints on $f_{\rm
NL}$ without relying on non-Gaussian numerical simulations. The
perturbation approach works when the skewness parameters multiplied by
variance, $S^{a}\sigma_0$, is much smaller than unity, i.e., $|f_{\rm
NL}|\ll 3300$ for CMB and $|f_{\rm NL}|\ll 5000$ for LSS, both of which
are satisfied by the current observational constraints from WMAP
\citep{Komatsu2003,Creminelli2005,Spergel2006}. We have shown that the
perturbation predictions agree with non-Gaussian numerical realizations
very well.   

We have used the Fisher matrix analysis to estimate the projected
constraints on $f_{\rm NL}$ expected from the observations of CMB and
LSS. We have found that the projected 1-$\sigma$ error on $f_{\rm NL}$
from the WMAP should reach 50, which is consistent with the MF
analysis given in \cite{Komatsu2003,Spergel2006}, and is comparable to
the current constraints from the bispectrum analysis given in
\cite{Komatsu2003,Creminelli2005,Spergel2006}. 
The MFs from the WMAP 8-year and Planck observations should be sensitive to
$|f_{\rm NL}|\sim 40$ and 20, respectively, at the 68\% confidence level.

As the MFs are solely determined by the weighted sum of the bispectrum
for $|f_{\rm NL}|\ll 3300$ for CMB and $|f_{\rm NL}|\ll 5000$ for LSS,
the MFs do not contain information more than the bispectrum. However,
this does not imply that the MFs are useless for measuring primordial
non-Gaussianity by any means. The important distinction between the MFs
and bispectrum is that the MFs are intrinsically defined in real space,
while the bispectrum is defined in  Fourier space. The systematics in
the data are most easily dealt with in real space, and thus the MFs
should be quite useful in this regard. Therefore, in the presence of
real-world issues such as inhomogeneous noise, foreground, masks, etc.,
these two approaches should be used to check for consistency of the
results.  

In this paper we have calculated the MFs from primordial
non-Gaussianity with a scale-independent $f_{\rm NL}$. It is easy to
extend our calculations to a scale-dependent $f_{\rm NL}$. All one
needs to do is to calculate the form of the bispectrum with a
scale-dependent $f_{\rm NL}$ \citep[e.g.,][]{Babich2004,Liguori2006},
and use it to obtain the skewness parameters, $S^{(a)}$
(Eqs.~[\ref{eq:scmb0}--\ref{eq:scmb2}] for CMB and
Eqs.~[\ref{eq:s0_lss}--\ref{eq:s2_lss}] for LSS). The MFs are then
given by equation~(\ref{eq:MFs_perturb}) in terms of the skewness
parameters.  Also, we have not included non-Gaussianity from secondary
anisotropy such as the Sunyaev-Zel'dovich effect, Rees-Sciama effect,
patchy reionization, weak lensing effect, extragalactic radio sources,
etc. It is again straightforward to calculate the MFs from these
sources using our formalism, as long as the form of the bispectrum is
known \citep[e.g.,][]{SG1999,GS1999,CH2000,KS2001,VS2002}.

The constraints on primordial non-Gaussianity from the MFs of LSS in a
galaxy survey covering 1~$h^{-3}~{\rm Gpc}^3$ volume are about 5 times
weaker than those from the MFs of CMB in the WMAP data. One would
therefore need the survey volume as large as 25~$h^{-3}~{\rm Gpc}^3$
to make the LSS constraints comparable to the WMAP constraints (using
the MFs only). This could be done by a survey covering $\sim
2000$~deg$^2$ at $3.5<z<6.5$. One would obviously need more volume to
make it comparable to the Planck data. The MFs from LSS are less
sensitive to primordial non-Gaussianity because non-Gaussianity from
the non-linear evolution of gravitational clustering exceeds the
primordial contribution at $R<200$, $120$, and 80~$h^{-1}$~Mpc at $z=0$,
2, and 5, respectively, which severely limits the amount of LSS data
available to constrain primordial non-Gaussianity.

\acknowledgments
C.~H. acknowledges support from a JSPS (Japan Society for the
Promotion of Science) fellowship. T.~M. acknowledges the support from
the Ministry of Education, Culture, Sports, Science, and Technology,
Grant-in-Aid for Encouragement of Young Scientists (No. 15740151).
E.~K. acknowledges support from an Alfred P. Sloan Research Fellowship.


\appendix
\section{Measuring Minkowski Functionals from CMB and LSS}
\label{app:sim}
In this Appendix we describe our methods for computing the MFs from the
CMB and LSS data. We also describe our simulations of CMB and LSS.

\subsection{Computational Method: CMB}\label{app:comp_cmb}
We estimate the MFs from pixelized CMB sky maps by integrating a
combination of first and second angular derivatives of temperature
anisotropy, ${\cal I}_k$, over the sky \citep{SG1998}, 
\begin{equation}
\label{eq:mfcomp}
V_k^{(2)}(\nu)=\frac{\sum_{i=1}^{n_{\rm pix}}w_i{\cal I}_k({\mathbf
 \Omega}_i)}{\sum_{i=1}^{n_{\rm pix}}w_i},
\end{equation}
where ${\mathbf \Omega}_i$ is the unit vector pointing toward a given
position on the sky. We set the weight of $i-$th pixel, $w_i$, to be 1
when the pixel at ${\mathbf \Omega}_i$ is outside of the survey mask,
and 0 otherwise. We calculate ${\cal I}_k$ at ${\mathbf \Omega}_i$ from
covariant derivatives of temperature anisotropy divided by its standard
deviation, $u({\mathbf \Omega}_i)\equiv (\Delta T/T)/\sigma_0$, 
\begin{eqnarray}
{\cal I}_0&=&\Theta(u-\nu), \\
{\cal I}_1&=&\frac{1}{4}F(u-\nu)\sqrt{u^2_{;\theta}+u^2_{;\phi}}, \\
{\cal I}_2&=&\frac{1}{2\pi}F(u-\nu)\frac{2u_{;\theta}u_{;\phi}
u_{;\theta\phi}-u^2_{;\theta}u_{;\phi\phi}-
u^2_{;\phi}u_{;\theta\theta}}{u^2_{;\theta}+u^2_{;\phi}}.
\end{eqnarray}
The function $F(u-\nu)$ has a value of $1/\Delta\nu$ ($\Delta\nu$ is the
binning width of $\nu$) when $u$ is within
$[\nu-\Delta\nu/2,\nu+\Delta\nu/2]$, and 0 otherwise. Covariant
derivatives are related to the partial derivatives as 
\begin{eqnarray}
u_{;\theta}&=&u_{,\theta}, \\
u_{;\phi}&=&\frac{1}{\sin\theta}u_{,\phi}, \\
u_{;\theta\theta}&=&u_{,\theta\theta}, \\
u_{;\theta\phi}&=&\frac{1}{\sin\theta}u_{,\theta\phi}-
\frac{\cos\theta}{\sin^2\theta}u_{,\phi}, \\
u_{;\phi\phi}&=&\frac{1}{\sin^2\theta}u_{,\phi\phi}+
\frac{\cos\theta}{\sin\theta}u_{,\theta}.
\end{eqnarray}

The first derivative of the temperature field $u_{,i}(i=\theta,$ or
$\phi)$ are calculated in Fourier space as; 
\begin{eqnarray}
u_{,i}&=&\sum_{lm}a_{lm}Y_{lm,i} \\
Y_{lm,\theta}&=&\left\{
\begin{array}{lc}
\displaystyle
\frac{ l}{ \tan\theta}Y_{lm}-\sqrt{\frac{ 2l+1}{ 2l-1}(l^2-m^2)}\frac{
1}{ \sin\theta}Y_{l-1 m} & (|m|<l) \\ 
\displaystyle
\frac{l}{\tan\theta}Y_{lm} & (m=l)
\end{array}
\right. \\
Y_{lm,\phi}&=&imY_{lm} \\
\end{eqnarray}
The second derivatives are also computed as
\begin{eqnarray}
u_{,ij}&=&\sum_{lm}a_{lm}Y_{lm,ij} \\
Y_{lm,\theta\theta}&=&\left[-\left(l(l+1)-\frac{m^2}{\sin^2\theta}\right)
Y_{lm}-\frac{1}{\tan\theta}Y_{lm,\theta}\right] \\
Y_{lm,\theta\phi}&=&imY_{lm,\theta} \\
Y_{lm,\phi\phi}&=&-m^2Y_{lm}
\end{eqnarray}
We calculate the MFs of temperature anisotropy for $\nu$ from $-3$ to
$3$ with $\Delta_\nu=0.24$, which yields $25$ bins per each MF. 

\subsection{Simulating WMAP Data}\label{app:sim_wmap}
In order to quantify uncertainties in the estimated MFs from cosmic
variance and various effects, we simulate Gaussian temperature
anisotropy maps with noise characteristics of the WMAP data. We
generate 1000 Gaussian realizations of CMB temperature anisotropy
using the HEALPix package \citep{Gorski1998}. We use $N_{\rm
side}=128$, $256$ and $512$ for $\theta_s\ge 40$, $40>\theta_s\ge 10$,
$10>\theta_s$ respectively. The number of pixels is given by $n_{\rm
pix}=12N_{\rm side}^2$. From each sky realization we construct eight
simulated maps of WMAP differential assemblies (DAs), Q1, Q2, V1, V2,
W1, W2, W3, and W4, by convolving the sky map with the beam transfer
function in each DA \citep{WMAPbeam}, and adding independent Gaussian
noise realizations following the noise pattern in each DA. Each pixel
is given noise variance of $\sigma_{0,{\rm noise}}^2/N_{\rm
obs}({\mathbf \Omega}_i)$, where $N_{\rm obs}({\mathbf \Omega}_i)$ is
the number of observations per pixel, and $\sigma_{0,{\rm noise}}$ is
given in \citet{WMAPsummary}. We then co-add the eight maps by
weighting each map by $\bar{N}_{\rm obs}/\sigma_{0,{\rm noise}}^2$,
where $\bar{N}_{\rm obs}$ is the full-sky average of $N_{\rm
obs}({\mathbf \Omega}_i)$. Finally, we mask the co-added map by the
$Kp0$ Galaxy mask (including point-source mask) provided by
\citet{WMAPfg}. This mask leaves $76.8$\% of the sky available for the
subsequent data analysis. In addition to WMAP one-year data, we also
simulate the future WMAP eight-year data, by simply multiplying
$N_{\rm obs}$ by a factor of $8$.

Before we estimate the MFs from each simulated map, we smooth it using a
Gaussian filter with a smoothing scale of $\theta_s$,
\begin{equation}
\label{eq:gaussfil}
W_l=\exp\left[-\frac12l(l+1)\theta_s^2\right].
\end{equation}
To remove the effect of survey mask, we calculate the Minkowski
Functionals by limiting the pixels where the five measurements,
$\sigma, \sigma_1,$ and $S^{(k)}(k=0, 1, \& 2)$, have nearly equal
values between the field with survey mask and that without survey
mask.  We use the pixels where the difference is within
$5\%$ of each standard deviation for $\theta_s= 20$ and $40$ arcmin
and $10\%$ for $\theta_s=70$ and $100$ arcmin. For $\theta_s=5$ and
$10$ arcmin, we use the pixels as long as they are away from
the boundary of the mask by more than $2\theta_s$.
Table \ref{tab:skyfrac}
lists the sky fraction used in the analysis for each smoothing scale.
\begin{table*}[t]
\caption{Sky fraction used for the calculation of MFs}
\begin{center}
\begin{tabular}{cc}
  \hline\hline
 $\theta_s$ & sky fraction [\%]\\ 
  \hline
  $100$ & $13$  \\
   $70$ & $26$  \\
   $40$ & $41$  \\
   $20$ & $62$  \\
   $10$ & $73$  \\ 
   $ 5$ & $75$  \\ \hline
\end{tabular}
\end{center}
\label{tab:skyfrac}
\end{table*}
The mean density over the pixels which are used for the calculation of
MFs is not completely zero and thus we subtract it from the density
field to satisfy zero mean in each realization.

\subsection{Simulating Planck Data}\label{app:sim_planck}

For the simulations of the Planck data, we follow the same procedure
as the WMAP simulations. Each realization is a coadded map of 9 bands
with the inverse weight of the noise variance listed in Table
\ref{tab:planck}. We approximate the beam transfer function as a
Gaussian function, $\exp(-\theta_{\rm beam}^2l(l+1)/2)$, where
$\theta_{\rm beam}=\theta_{\rm FWHM}/\sqrt{8\ln(2)}$.  We add the
homogeneous noise distribution with the noise variance per pixel
$\sigma_{\rm noise}^2$ given by
\begin{equation}
\frac{1}{\sigma_{\rm noise}^2}=\sum_i \left(\frac{T}{\Delta
T}\right)^{(i)2} \frac{4\pi/n_{\rm pix}}{(\theta_{\rm
FWHM}^{(i)})^2},
\end{equation}
where $i$ denotes each band, and $n_{\rm pix}=12N_{\rm side}^2$ is the
number of pixels in simulated maps. We use the $Kp0$ mask to define the
survey area for the Planck simulations, in exactly the same manner as
for the WMAP simulations.

\begin{table*}[t]
\caption{%
 Approximate instrument specifications of the Planck
 satellite ({\sf http://www.rssd.esa.int/Planck}). The central
 frequency, full width at half maximum (FWHM) of beam size, $\theta_{\rm
 FWHM}$, and noise per pixel ($\Delta T/T$ per $\theta_{\rm FWHM}^2$)
 for 14 months of observations are shown per each frequency channel.
 }%
\begin{center}
\begin{tabular}{ccccccccccc}
  \hline\hline
 Instrument & \multicolumn{3}{c}{LFI} & &\multicolumn{6}{c}{HFI} \\
  \hline
 Center Frequency [GHz] & $30$ & $44$ & $70$ & & $100$ & $143$ & $217$ & $353$ & $545$ & $857$ \\ 
 Beam size $\theta_{\rm FWHM}$ [arcmin] & $33$ & $24$ & $14$ & & $9.5$ & $7.1$ & $5$ & $5$ & $5$ & $5$ \\
 Pixel Noise $\Delta T/T$ [$10^6$] & $2$ & $2.7$ & $4.7$ & & $2.5$ & $2.2$ & $4.8$ & $14.7$ & $147$ & $6700$ \\
  \hline
\end{tabular}
\end{center}
\label{tab:planck}
\end{table*}

\subsection{Computational Method: LSS}\label{app:comp_lss}
We use two complementary routines to compute the MFs of a density field on
the grids. The first approach is often called {\it Koenderink
invariants} \citep{Koenderink1984} in which the surface integrals of the
curvature are transformed into the volume integral of invariants formed from the
first and second derivatives of the density fluctuations. The second
method, which is called {\it Crofton's formula}
\citep{Crofton1868,SB1997, Koenderink1984}, is based on the integral
geometry and the calculation reduce to simply counting the elementary
cells (e.g., cubes, squares, lines, and points for the cubic meshes).
The outline of these methods are summarized in \cite{SB1997} and the
observational application to SDSS galaxy samples are performed by
\cite{Hikage2003}. 

\subsection{Simulating LSS Data with Primordial Non-Gaussianity}
\label{app:sim_lss}
We calculate the MFs of density field in a cubic box with a length of
$1~h^{-1}$~Gpc, but ignore the observational effects such as survey
geometry, for simplicity. We also ignore non-linear gravitational
clustering or galaxy bias in order to isolate the effect from primordial
non-Gaussianity. (The purpose of this simulation is to check the
accuracy of our perturbation predictions for the form of MFs from
primordial non-Gaussianity.) To simulate the LSS data with primordial
non-Gaussianity, we first generate a Gaussian potential field in a cubic
box with a length of $1~h^{-1}$~Gpc, assuming that the power spectrum of
potential is $P_\phi(k)\propto k^{n_s-4}$ where $n_s=0.967$. We
inversely Fourier-transform it into real space to obtain $\phi({\mathbf
x})$. (The number of grids is $128^3$.) We then construct a non-Gaussian
potential field, $\Phi({\mathbf x})$, using
equation~(eq.[\ref{eq:ngpotential2}]) for a given $f_{\rm NL}$. We
finally convert it to the matter density field by multiplying $\Phi$ by
$M(k)$ in Fourier space (see Eq.~[\ref{eq:Mk}]). We have generated 2,000
realizations of the non-Gaussian density field. 

\section{Derivation of Galaxy Bispectrum}\label{app:Bg}
In this Appendix we derive the perturbative formula for the galaxy
bispectrum including primordial non-Gaussianity, non-linearity in
gravitational clustering, and non-linearity in galaxy biasing, in the
weakly non-linear regime \citep{Verde2000,Scocci2004}. 
In the weakly non-linear regime, it would be reasonable to assume that
the galaxy biasing is local and deterministic. We then expand the galaxy
density contrast, $\delta_{\rm g}$, perturbatively in terms of the
underlying matter density contrast, $\delta_{\rm m}$, as \citep{FG1993} 
\begin{equation}
\delta_{\rm g}(z)=b_0(z)+b_1(z)\delta_{\rm m}(z)
+\frac{b_2(z)}{2}\delta_{\rm m}^2(z)
+{\cal O}(\delta_{\rm m}^3).
\label{eq:fg93}
\end{equation}
where $b_0(z)$ is determined such that $\langle \delta_{\rm
g}(z)\rangle=0$. Here, $b_1(z)$ and $b_2(z)$ are the time-dependent
galaxy bias parameters. The power spectrum and bispectrum of the galaxy
distribution, $P_{\rm g}$ and $B_{\rm g}$, respectively, are then given
by those of the underlying matter distribution as 
\begin{eqnarray}
P_{\rm g}(k,z)&=&b_1^2(z)P_{\rm m}(k,z), \\
B_{\rm g}(k_1,k_2,k_3,z)&=&b_1^3(z)B_{\rm m}(k_1,k_2,k_3,z)
+b_1^2(z)b_2(z)[P_{\rm m}(k_1,z)P_{\rm m}(k_2,z)+(\mbox{cyc.})],
\end{eqnarray}
respectively. If the underlying mass distribution obeyed Gaussian
statistics, its bispectrum would vanish exactly, $B_{\rm m}\equiv 0$;
however, the non-linear evolution of density fluctuations due to
gravitational instability makes $\delta_{\rm m}$ slightly non-Gaussian
in the weakly non-Gaussian regime, yielding non-zero bispectrum. 

The second-order correction to the density fluctuations from non-linear
gravitational clustering gives the following equation,
\begin{equation}
\tilde{\delta}_{{\rm m},{\mathbf k}}(z)=
\tilde{\delta}_{{\rm L},{\mathbf k}}(z)
+\int d^3qF_2({\mathbf
q},{\mathbf k}-{\mathbf q}) \tilde{\delta}_{{\rm L},{\mathbf q}}(z)
\tilde{\delta}_{{\rm L},{\mathbf k}-{\mathbf q}}(z),
\end{equation}
where $\tilde{\delta}_{{\rm L},{\mathbf k}}(z)$ is the {\it linear} (but
non-Gaussian) density fluctuations, and  
\begin{equation}
\label{eq:f2}
F_2({\mathbf k}_1,{\mathbf k}_2)=\frac{5}{7}
+\frac{{\mathbf k}_1\cdot{\mathbf k}_2}{
2k_1k_2}\left(\frac{k_1}{k_2}+\frac{k_2}{k_1}\right)+\frac{2}{7}
\frac{({\mathbf k}_1\cdot{\mathbf k}_2)^2}{k_1^2k_2^2},
\end{equation}
is the time-independent kernel describing mode-coupling due to
non-linear clustering of matter density fluctuations in the weakly
non-linear regime. Equation~(\ref{eq:f2}) is exact only in an
Einstein-de Sitter universe, but the corrections in other cosmological
models are small \citep[e.g.,][]{Bernardeau1994}. The power spectrum and
bispectrum of the underlying  mass density distribution, $P_{\rm m}$ and
$B_{\rm m}$, are thus given in terms of the linear and non-linear
contributions: 
\begin{eqnarray}
P_{\rm m}(k,z)&=&P_{\rm L}(k,z), \\
B_{\rm m}(k_1,k_2,k_3,z)
&=&
B_{\rm prim}(k_1,k_2,k_3,z) 
+ 2\left[F_2({\mathbf k}_1,{\mathbf k}_2)P_{\rm L}(k_1,z) 
P_{\rm L}(k_2,z)
+(\mbox{cyc.})\right].
\end{eqnarray}
Note that we have ignored the non-linear contributions in the power
spectrum. That is to say, the power spectrum is still described by
linear perturbation theory.

The remaining task is to relate $\delta_{\rm L}(z)$ to Bardeen's
curvature perturbations during the matter era, $\Phi$. One may use
Poisson's equation for doing this: 
\begin{equation}
 k^2\tilde{\Phi}_{\mathbf k}T(k)
=4\pi G\rho_{\rm m}(z)\frac{\tilde{\delta}_{{\rm m},{\mathbf k}}(z)}{(1+z)^2}
=\frac32\Omega_{\rm m}H_0^2\tilde{\delta}_{{\rm m},{\mathbf k}}(z)(1+z),
\end{equation}
where $T(k)$ is the linear transfer function that describes the
evolution of density fluctuations during the radiation era and the
interactions between photons and baryons \citep{EH1999}. Note that
$\Phi$ is independent of time during the matter era. At very early
times, say, $z=z_*\sim 10^3$, the non-linear evolution may be safely
ignored at the scales of interest, and one obtains
\begin{equation}
 \tilde{\delta}_{{\rm L},{\mathbf k}}(z_*)
= \frac{M(k)\tilde{\Phi}_{\mathbf k}}{1+z_*},
\end{equation}
where
\begin{equation}
 M(k)\equiv \frac23\frac{k^2T(k)}{\Omega_{\rm m}H_0^2}.
\label{eq:Mk}
\end{equation}
Therefore, using the quadratic non-Gaussian model given in
equation~(\ref{eq:ngpotential2}), one obtains the linear power spectrum
at $z=z_*$, 
\begin{equation}
\label{eq:pphi}
P_{\rm L}(k,z_*)=\frac{M^2(k)}{(1+z_*)^2}
P_\phi(k),
\end{equation}
and the primordial bispectrum at $z=z_*$, 
\begin{eqnarray}
\nonumber
B_{\rm prim}(k_1,k_2,k_3,z_*)&=&2f_{\rm NL}
\frac{M(k_1)M(k_2)M(k_3)}{(1+z_*)^3}[P_\phi(k_1)P_\phi(k_2)+(\mbox{cyc.})]\\ 
&=&
2f_{\rm NL}(1+z_*)
\left[\frac{P_{\rm L}(k_1,z_*)P_{\rm L}(k_2,z_*)M(k_3)}{M(k_1)M(k_2)}+
(\mbox{cyc.})\right],
\end{eqnarray}
where $P_\phi(k)\propto k^{n_s-4}$ is the power spectrum of $\phi$ and
we have ignored the higher-order terms. We then use the linear growth
rate of density fluctuations, $\delta_{\rm L}\propto D(z)$, to evolve
the linear bispectrum forward in time:
\begin{equation}
B_{\rm prim}(k_1,k_2,k_3,z)=
2f_{\rm NL}\frac{(1+z_*)D(z_*)}{D(z)}
\left[\frac{P_{\rm L}(k_1,z)P_{\rm L}(k_2,z)M(k_3)}{M(k_1)M(k_2)}+
(\mbox{cyc.})\right].
\label{eq:bl}
\end{equation}
One may simplify this expression by normalizing the growth rate such
that 
\begin{equation}
 D(z_*)=\frac1{1+z_*}.
\end{equation}
Note that this condition gives the normalization of $D(z)$ that is
actually independent of the choice of $z_*$, when $z_*$ is taken to be
during the matter era. 

By putting all the terms together, we finally obtain  the following form
of $B_{\rm g}(k_1,k_2,k_3,z)$: 
\begin{eqnarray}
\nonumber
B_{\rm g}(k_1,k_2,k_3,z)
&=& 2f_{\rm NL}\frac{b_1^3(z)}{D(z)}
\left[\frac{P_{\rm m}(k_1,z)P_{\rm m}(k_2,z)M(k_3)}{M(k_1)M(k_2)}+
(\mbox{cyc.})\right] \\
\nonumber
& &+ 2b_1^3(z)\left[F_2({\mathbf k}_1,{\mathbf k}_2)P_{\rm m}(k_1,z) 
P_{\rm m}(k_2,z)
+(\mbox{cyc.})\right]\\
& &+b_1^2(z)b_2(z)[P_{\rm m}(k_1,z)P_{\rm m}(k_2,z)+(\mbox{cyc.})].
\label{eq:Bgfinal}
\end{eqnarray}
We use this formula to calculate the skewness parameters that are used
for the MFs of the galaxy distribution. 

While the perturbative formula for the MFs derived by
\citet{Matsubara2003} (See \S~\ref{sec:pb_general}) works well for
$\Phi$, it is not immediately clear if it works for $\delta_{\rm g}$
because of the $k$-dependent coefficient, $\tilde{\delta}_{{\rm
g},{\mathbf k}} \propto M(k)\tilde{\Phi}_{\mathbf k}$. In
Appendix~\ref{app:poten} we show that the perturbative formula still
works, as long as $f_{\rm NL}$ is not very large. The current
observational constraints on $f_{\rm NL}$ already guarantee that the
perturbative formula for the MFs of the galaxy distribution provides an
excellent approximation. 

\section{MFs of CMB: Analytical Formula vs Simulations in the
 Sachs--Wolfe Limit}
\label{app:cmbsim} 
We compare the perturbative formula of MFs for CMB with Monte-Carlo
realizations of non-Gaussian temperature anisotropy in the Sachs-Wolfe
regime, in order to check the accuracy of our formalism.

\begin{figure*}[t]
\begin{center}
\includegraphics[width=12cm]{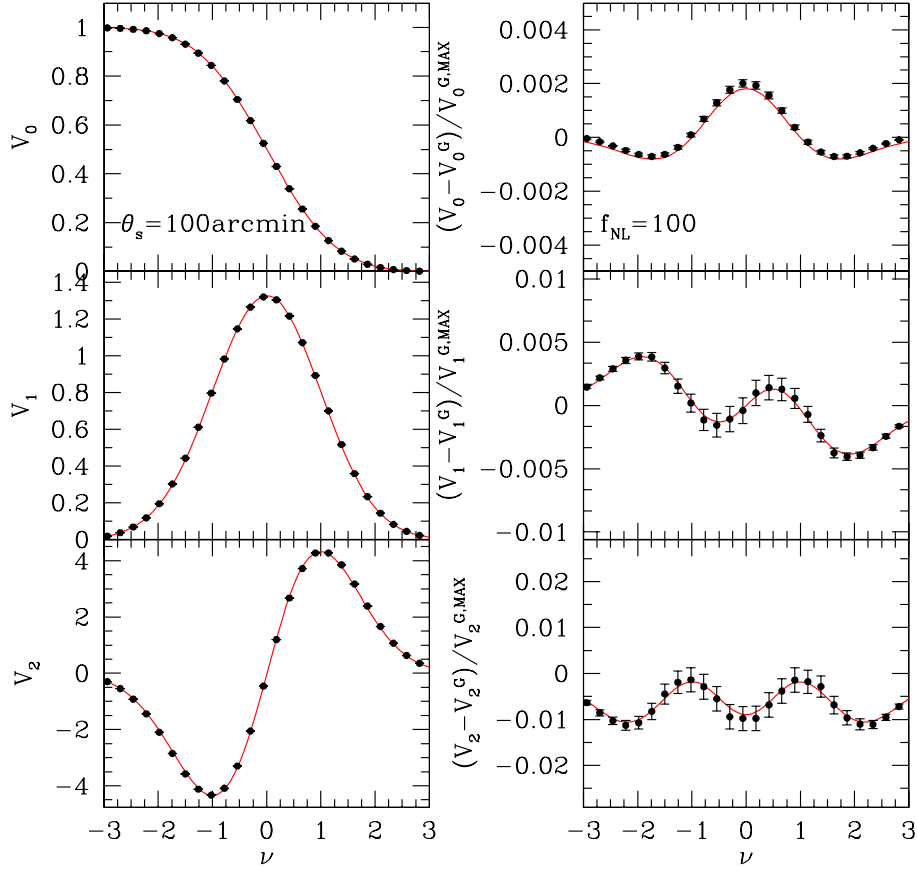}
\caption{%
  Comparison between the MFs calculated from the analytical perturbation
 predictions (solid lines) and the numerical simulations (symbols) in
 the Sachs-Wolfe limit. We have used $\theta_s=100$~arcmin and $f_{\rm
 NL}=100$. 
 }
\label{fig:cmbmf_sw}
\end{center}
\end{figure*}

The angular power spectrum spectrum is set to be $l(l+1)C_l^{\rm
SW}/2\pi=10^{-10}$. In the Sachs-Wolfe limit, $\Delta T^{\rm SW}/T=-\Phi/3$, the
non-Gaussian maps of CMB temperature anisotropy may be constructed from
the Gaussian maps, $\Delta T_G/T$, by the following simple mapping:
\begin{equation}
\frac{\Delta T^{\rm SW}}{T}=\frac{\Delta T_G}{T}-3f_{\rm NL}
\left[\left(\frac{\Delta T_G}{T}\right)^2-\left\langle\left(\frac{\Delta 
T_G}{T}\right)^2\right\rangle\right].
\end{equation}
We calculate the MFs from 6000 realizations of the non-Gaussian CMB
maps and compare them with the perturbation predictions
(Eq.[\ref{eq:MFs_perturb}]). The skewness parameters can be calculated
from the reduced bispectrum of CMB in the Sachs-Wolfe limit,
\begin{equation}
b_{l_1l_2l_3}=-6f_{\rm NL}(C_{l_1}^{\rm SW}C_{l_2}^{\rm SW}+
C_{l_2}^{\rm SW}C_{l_3}^{\rm SW}+C_{l_3}^{\rm SW}C_{l_1}^{\rm SW}).
\end{equation}
Figure~\ref{fig:cmbmf_sw} shows that the MFs from Monte-Carlo
realizations agree with the perturbation predictions very well.
The comparison with the full simulations that include the full radiation
transfer function will be reported elsewhere.
(For subtlety in this comparison arising from pixelization and boundary
effects, see \S~\ref{sec:method_lss}.)

\section{Validity of Perturbative Formulae for $N-$th order corrections of Primordial Non-Gaussianity}
\label{app:poten}

We consider the primordial non-Gaussianity extended to $n-$th order
corrections of a primordial potential field; 
\begin{equation}
\label{eq:ngpotentialn}
\Phi=\epsilon\phi_0+\frac{\epsilon^2 f_{\rm NL}^{(1)}}{2!}
(\phi_0^2-\langle\phi_0^2\rangle)+
\frac{\epsilon^3f_{\rm NL}^{(2)}}{3!}\phi_0^3+\frac{\epsilon^4f_{\rm
NL}^{(3)}}{4!}(\phi_0^4-\langle\phi_0^4\rangle)+ \cdot\cdot\cdot 
+\frac{\epsilon^nf_{\rm NL}^{(n-1)}}{n!}
(\phi_0^n-\langle\phi_0^n\rangle)+\cdot\cdot\cdot,
\end{equation}
where $\epsilon\phi_0$ is an auxiliary random-Gaussian field and
$f_{\rm NL}^{(n)}$ represents the coefficients of $n$-th order of
$\phi$.  For convenience, we separate the random-Gaussian field into
two parts; the $\epsilon$ represents the amplitude of the primordial
potential power spectrum which has the order of $10^{-4}$ and thereby
the fluctuation of $\phi_0$ is the order unity.  The parameter $f_{\rm
NL}$ in the equation (\ref{eq:ngpotential2}) corresponds to $f_{\rm
NL}^{(1)}/2$.

The Fourier transform of $\Phi$ is written by
\begin{equation}
\tilde{\Phi}=\epsilon\tilde{\phi}_0
+\frac{\epsilon^2 f_{\rm NL}^{(1)}}{2!}\tilde{\Phi}_0^{(2)}
+\cdot\cdot\cdot
+\frac{\epsilon^nf_{\rm NL}^{(n-1)}}{n!}\tilde{\Phi}_0^{(n)}
+\cdot\cdot\cdot
\end{equation}
where $\tilde{\Phi}_0^{(n)}({\mathbf k})$ is the Fourier transform
of the $n$-th order term, $\phi_0^n-\langle\phi_0^n\rangle$, given by
\begin{equation}
\tilde{\Phi}_0^{(n)}({\mathbf k})=\frac{1}{(2\pi)^{3n-3}}
\int d{\mathbf k}_1\cdot\cdot\cdot \int d{\mathbf k}_{n-1}
\tilde{\phi}_0^\ast({\mathbf k}_1)\cdot\cdot\cdot 
\tilde{\phi}_0^\ast({\mathbf k}_{n-1})\tilde{\phi}_0({\mathbf k}+
{\mathbf k}_1+\cdot\cdot\cdot{\mathbf k}_{n-1}) 
-(2\pi)^3\delta_D({\mathbf k})\langle \phi_0^n \rangle
\end{equation}
The polyspectra of $\Phi$, $P^{(n)}_\Phi$ $(n\ge 2)$, are defined by
\begin{equation}
\langle \Phi({\mathbf k}_1)\cdot\cdot\cdot 
\Phi({\mathbf k}_n)\rangle_c
\equiv(2\pi)^3\delta_D({\mathbf k}_1+\cdot\cdot\cdot {\mathbf k}_n)
P_\Phi^{(n)}({\mathbf k}_1,\cdot\cdot\cdot,{\mathbf k}_{n})
\end{equation}
where $P^{(2)}_\Phi$ and $P^{(3)}_\Phi$ correspond to the power
spectrum and the bispectrum of $\Phi$ respectively.

According to the diagrammatic method by \citet{Matsubara1995}, the
lowest order terms of $\epsilon$ for the connected part of $n$-th order
moment, called {\it cumulants}, are the $n-1$ products of the quadratic
moment as follows;
\begin{eqnarray}
P_\Phi^{(2)}&=&\epsilon^2 P_{\phi_0} \\
P_\Phi^{(3)}&=&\epsilon^4 f_{\rm NL}^{(1)}
(P_{\phi_0}(k_1)P_{\phi_0}(k_2)+(\mbox{cyc.})) \\
P_\Phi^{(4)}&=&\epsilon^6 f_{\rm NL}^{(1)2}(P_{\phi_0}(k_1)P_{\phi_0}(k_2)
P_{\phi_0}(|{\mathbf k}_1+{\mathbf k}_3|)+(\mbox{sym.})(12)) \nonumber \\
&+&\epsilon^6 f_{\rm NL}^{(2)}(P_{\phi_0}(k_1)P_{\phi_0}(k_2)
P_{\phi_0}(k_3)+(\mbox{cyc.})) \\
P_\Phi^{(5)}&=&\epsilon^8 f_{\rm NL}^{(1)3}(P_{\phi_0}(k_1)P_{\phi_0}(k_2)
P_{\phi_0}(|{\mathbf k}_1+{\mathbf k}_3|)
P_{\phi_0}(|{\mathbf k}_2+{\mathbf k}_4|)+(\mbox{sym.})(60)) \nonumber \\
&+&\epsilon^8 f_{\rm NL}^{(1)}f_{\rm NL}^{(2)}
(P_{\phi_0}(k_1)P_{\phi_0}(k_2)P_{\phi_0}(k_3)
P_{\phi_0}(|{\mathbf k}_1+{\mathbf k}_4|)+(\mbox{sym.})(60)) \nonumber \\
&+&\epsilon^8 f_{\rm NL}^{(3)}(P_{\phi_0}(k_1)P_{\phi_0}(k_2)
P_{\phi_0}(k_3)P_{\phi_0}(k_4)+(\mbox{cyc.})) \\
&\cdot\cdot\cdot& \nonumber \\
P_\Phi^{(n)}&=&\epsilon^{2n-2}\sum^{d_1+2d_2+\cdot\cdot\cdot
nd_n=n-2}_{d_1,d_2,...,d_n}\left(\prod_{m=1,...,n}f_{\rm
NL}^{(m)~d_m}\right)\left[\prod_{k_{AB}(d_1,d_2,...,d_n)}^{n-1}
P_{\phi_0}(|{\mathbf k}_A+{\mathbf k}_B|) + (\mbox{sym.})\right]
\end{eqnarray}
where (sym.)($n$) means the addition of $n$ terms with the subscripts
symmetric to the previous term and the edge $(AB)$ is one of the edges
in a tree graph $(d_1,d_2,...,d_n)$ which satisfies the condition that
$d_1+2d_2+\cdot\cdot\cdot +nd_n=n-2$.

We obtain the $n-$th polyspectrum of $\tilde{\delta}_{\rm L,{\mathbf
k}}$, $P^{(n)}_{\rm L}$, by
\begin{equation}
P^{(n)}_{\rm L}({\mathbf k}_1,\cdot\cdot\cdot,{\mathbf k}_{n-1})
=M(k_1)M(k_2)\cdot\cdot\cdot M(k_n)P^{(n)}_\Phi.
\end{equation}

The $n$-th order terms of the perturbative formula (eq.[1]) 
are represented by
\begin{equation}
\label{eq:pb_norder}
\frac{\langle \delta_{\rm L}^n \rangle_c}{\langle \delta_{\rm L}^2
\rangle^{n/2}},~ \frac{\langle \delta_{\rm L}^{n-1}\nabla^2\delta_{\rm
L} \rangle_c}{\langle \delta_{\rm L}^2 \rangle^{n/2-1} \langle (\nabla
\delta_{\rm L})\cdot(\nabla \delta_{\rm L})\rangle},~ \frac{\langle
\delta_{\rm L}^{n-3}(\nabla \delta_{\rm L}\cdot\nabla \delta_{\rm
L})\nabla^2\delta_{\rm L} \rangle_c}{ \langle \delta_{\rm L}^2
\rangle^{n/2-2}\langle (\nabla \delta_{\rm L})\cdot(\nabla \delta_{\rm
L})\rangle^2},...
\end{equation}
These terms are $n$-th order cumulants of the products of $\delta_{\rm
L}, \nabla \delta_{\rm L}$, and $\nabla^2\delta_{\rm L}$ divided by
$n/2$ times product of the corresponding combination of the second
moments $\langle \delta_{\rm L}^2 \rangle$, $\langle (\nabla
\delta_{\rm L})\cdot(\nabla \delta_{\rm L}) \rangle$.

The $n$-th order cumulants $\langle \delta_{\rm L}^n \rangle_c$ are
obtained by the inverse Fourier-transform of $P^{(n)}_{\rm L}$ as
\begin{eqnarray}
\langle \delta_{\rm L}^2 \rangle_c&=&\frac{1}{(2\pi)^3}\int d{\mathbf k}
P_\Phi^{(2)}(k)M(k)^2W(kR)^2 \\
\langle \delta_{\rm L}^{n} \rangle_c&=&\frac{1}{(2\pi)^{3n-3}}\int
d{\mathbf k}_1\cdot\cdot\cdot\int d{\mathbf k}_{n-1}
P_\Phi^{(n)}M(k_1)W(k_1R)\cdot\cdot\cdot M(k_n)W(k_nR)
\end{eqnarray}
The $n$-th order term of the perturbative formula (eq.[1])
has the following order;
\begin{equation}
\label{eq:primng_orderpara}
\frac{\langle \delta_{\rm L}^n \rangle_c}{\langle \delta_{\rm L}^2
\rangle_c^{n/2}} 
\sim \epsilon^{n-2}\sum^{d_1+2d_2+\cdot\cdot\cdot
nd_n=n}_{d_1,d_2,...,d_n}\left(\prod_{m=1,...,n}f_{\rm
NL}^{(m)~d_m}\right).
\end{equation}
The other $n$-th order terms in equation (\ref{eq:pb_norder})
have the same order of $\epsilon$ as 
$\langle \delta_{\rm L}^n \rangle_c/\langle \delta_{\rm L}^2 \rangle_c^{n/2}$.

The above equation is different from
the well-known hierarchical condition from the gravitational evolution;
\begin{equation}
\langle f^n \rangle_c\sim \langle f^2\rangle^{n-1}.
\end{equation}
Indeed, the skewness parameters due to the primordial non-Gaussianity
have a scale dependence of $M(k)$.  Nevertheless, the perturbation
works well as long as equation (\ref{eq:primng_orderpara}) is much
smaller than unity, which corresponds to
\begin{equation}
|f_{\rm NL}^{(n)}|\ll \epsilon^{-n}\simeq 10^{4n}
\end{equation}
Recent observations represented by WMAP (Komatsu et al. 2003) gave
constraints on $|f_{\rm NL}^{(1)}|<{\cal O}(10^2)$. Standard inflation
models predict that higher-order coefficients are the same order as
$f_{\rm NL}$ and thus the perturbation is applicable to the actual
primordial non-Gaussianity.
\end{document}